\documentclass{aa}

\usepackage{graphicx}
\usepackage[11pt]{moresize}
\usepackage{txfonts}
\usepackage[]{hyperref}
\usepackage{subcaption}
\usepackage{cleveref}
\usepackage{longtable}
\usepackage{xcolor}
\begin{document} 

   \title{Searching for central stars of planetary nebulae in \emph{Gaia} DR2\thanks{Table B.1 is only available in electronic form at the CDS via anonymous ftp to \url{cdsarc.u-strasbg.fr} (130.79.128.5) or via \url{http://cdsweb.u-strasbg.fr/cgi-bin/qcat?J/A+A/}}}

   \author{N. Chornay \and N. A. Walton}

   \institute{Institute of Astronomy, University of Cambridge, Madingley Road, Cambridge, CB3 0HA, United Kingdom\\
              \email{njc89@ast.cam.ac.uk}, \email{naw@ast.cam.ac.uk}}

   \date{Received January 23, 2020; accepted April 17, 2020}



  \abstract
   {Accurate distance measurements are fundamental to the study of Planetary Nebulae (PNe) but have long been elusive. The most accurate and model-independent distance measurements for galactic PNe come from the trigonometric parallaxes of their central stars, which were only available for a few tens of objects prior to the \emph{Gaia} mission.}
   {Accurate identification of PN central stars in the \emph{Gaia} source catalogues is a critical prerequisite for leveraging the unprecedented scope and precision of the trigonometric parallaxes measured by \emph{Gaia}. Our aim is to build a complete sample of PN central star detections with minimal contamination.}
   {We develop and apply an automated technique based on the likelihood ratio method to match candidate central stars in \emph{Gaia} Data Release 2 (DR2) to known PNe in the Hong Kong/AAO/Strasbourg H$\alpha$ (HASH) PN catalogue, taking into account the BP--RP colours of the \emph{Gaia} sources as well as their positional offsets from the nebula centres. These parameter distributions for both true central stars and background sources are inferred directly from the data.}
   {We present a catalogue of over 1000 \emph{Gaia} sources that our method has automatically identified as likely PN central stars. We demonstrate how the best matches enable us to trace nebula and central star evolution and to validate existing statistical distance scales, and discuss the prospects for further refinement of the matching based on additional data. We also compare the accuracy of our catalogue to that of previous works.}
   {}


   \authorrunning{N. Chornay \& N. A. Walton}
   \maketitle
%


\section{Introduction}

Planetary Nebulae (PNe) are an end stage of life for low and intermediate mass stars, a relatively short step on their evolutionary path after they depart from the tip of the Asymptotic Giant Branch (AGB) \citep{herwigAGB}. The star sheds its outer layers, growing brighter and hotter before ultimately cooling into a white dwarf. Ultraviolet light from the star ionises this rapidly expanding shell of gas, which reaches typical sizes on the order of light-years over the tens of thousands of years during which it is visible.

PNe are important in galactic evolution for their enrichment of the interstellar medium (ISM) with heavier elements \citep{johnsonnucleosynthesis, karakas_lattanzio_2014_nucleosynthesis}, joining other mechanisms of stellar mass loss such as supernovae. Their brightness and resulting visibility over large distances make PNe valuable chemical probes of not only the Milky Way but also nearby galaxies \citep{kwitterwhitepaper}. In addition, the Planetary Nebula Luminosity Function (PNLF) forms a useful rung in the cosmic distance ladder \citep{ciardullo2012pnlf}.

The number of PNe present in our galaxy at any given time is small relative to the stellar population on account of their short lifespans. However the set of 3500 or so confirmed and likely PNe that have been discovered only represents a fraction of those expected to be visible if all stars in a certain mass range go through the PN phase \citep{demarcobinaries2006}, with this inconsistency leaving open questions as to whether there are further requirements for PN formation, namely binary interactions \citep{jonesboffinnaturebinaries2017}. Understanding of PNe is limited in part by difficulties in constraining their distances \citep{smith2015}. Accurate distances are critical for meaningful astrophysical characterisation of PNe, from measuring physical sizes of individual PNe and the absolute magnitudes of their central stars to determining their lifetimes and formation rates.

The rapid evolution of Central Stars of PNe (CSPNe) generally prevents the application of usual methods of distance determination such as isochrone fitting. Thus a variety of distance measurement techniques have been developed, which fall into two broad categories (following \citet{frewsurfacebrightness2016}, henceforth FPB16).

Primary techniques measure the distances to individual PNe with varying degrees of accuracy and assumptions. Most involve modelling -- either of the nebula's expansion \citep{expansion2018}, of the environment (e.g. extinction distances, cluster or bulge membership, or location in external galaxies), or of the CSPN itself. The most direct primary distances measurements come from trigonometric parallaxes of CSPNe, but until recently these have only been available for few nearby objects with measurements from the United States Naval Observatory  (USNO) \citep{usno2007} and the Hubble Space Telescope (HST) \citep{hst2009}.

Secondary, or statistical distance scales, rely on finding a broadly applicable relationship that provides a means of estimating a physical parameter of the PN such as its physical size, given a distance-independent measurement, such as nebula surface brightness. The distance can then be determined from the relation of the physical parameter to a measured one, for example through comparison of physical to angular size, in a manner analogous to the distance modulus for stars with known absolute magnitudes. The determination of the relationships underlying such secondary methods require a calibrating set of objects whose distances are known independently.

Most PN distance estimates rely on secondary methods, but these methods are only as good as the quality and purity of the distances used to calibrate them. Incorrect distances or polluting objects can inflate errors well beyond the uncertainties stemming from measurement errors and intrinsic scatter. Thus, improved primary distances to a set of PNe provides a twofold benefit, as it betters not only the distances to that set of objects but also, through improved calibration of statistical distance scales, to the population as a whole.

Good primary distance measurements are rare. In their statistical distance scale calibration, FPB16 deemed only around 300 galactic and extragalactic PNe to have sufficiently reliable primary distances. The galactic selection they chose represented only around 5\% of confirmed galactic PNe, and their more relatively accurate primary distances were for extragalactic objects.

However, the situation is now changing with the recently launched \emph{Gaia} mission \citep{gaiamission}, which is conducting astrometric measurements -- positions, parallaxes, and proper motions -- of over a billion stars in the Milky Way, including many CSPNe. \citet{stanghellini2017} found a small number of CSPNe with parallax measurements in \emph{Gaia} DR1 \citep{gaiadr1}, while \citet{kimeswenger2018} found a larger sample in the most recent \emph{Gaia} data release, DR2 \citep{gaiadr2}, using a manual matching technique and a limited input catalogue. Most recently, \citet{gapn} and \citet{stanghellini2020} searched \emph{Gaia} DR2 using more complete input catalogues, but relied on a position-based matching approach that creates a risk of contamination.

Fully exploiting the data from \emph{Gaia} for the study of PNe requires a CSPN sample that is both complete and pure, which is time-consuming and difficult to achieve manually. As astronomical datasets become larger and are updated more frequently, automated techniques become increasingly useful, offering not only improved speed and consistency but also adaptability, making it easier to incorporate new data as it becomes available. In the case of \emph{Gaia}, future data releases will have more detections with improved photometry and parallaxes, so an automated technique will allow these to quickly be taken advantage of.

Our work aims to provide a more complete sample of CSPNe in \emph{Gaia} DR2 and to lay the groundwork for future data releases, through an automated matching process that we have developed that takes into account both relative position and colour information of \emph{Gaia} sources. In the remainder of this work we present the technique, the resulting catalogue of CSPNe in \emph{Gaia} DR2, a comparison of this catalogue to previous works, and finally some initial applications that use the subsets of this catalogue with the most accurate \emph{Gaia} parallaxes for astrophysical characterisation and distance scale evaluation.

\section{Methods} \label{methods}

Our starting point is the HASH PN catalogue\footnote{\url{https://hashpn.space}} from \citet{hashpn}. This catalogue represents the most complete catalogue of PNe available, containing at the time of writing\footnote{Version 4.6 of the HASH PN catalogue, downloaded on June 16, 2019, was used to produce the published catalogue for this work.} around 2500 spectroscopically confirmed PNe (following the criteria in \citet{pndiagnostics}) and 1000 possible and likely PNe, as well as objects that are commonly confused with PNe, such as HII regions, symbiotic stars, and reflection nebulae. HASH also collects together additional information about individual nebulae such as fluxes, angular sizes, and spectra; however it lacks structured positional data on known CSPNe.

PN catalogues have historically suffered from positional inaccuracies. Some uncertainty is intrinsic to PNe as extended objects: they have varied morphologies, and the full extents of their nebulae may not be visible, so different assessments of the nebula position are possible. Moreover in the cases where there is a known or apparent central star, some catalogues adopt this star's position as that of the PN, though the nebula and star positions can have significant offsets.

In the the Macquarie/AAO/Strasbourg H$\alpha$ PNe Catalogue (MASH) (\citet{mashpn}, \citet{mashpn2}), the precursor to HASH, the authors based their reported positions on the geometric centre of the visible nebula, and claimed uncertainties on the order of 1\arcsec to 2\arcsec. They found notable disagreements between their measured positions for known PNe and those in existing catalogues. This can in part be due to catalogue inhomogeneity noted above. However outright misidentification can also occur, particularly for compact PNe: the HASH authors found positions in the online SIMBAD database that were simply incorrect. We see in Sect. \ref{comparison} that such errors are still present.

HASH promises a homogeneous set of PN coordinates, primarily based on centroiding narrowband H$\alpha$ imagery of the PNe. This along with its completeness motivates our choice of it as an input catalogue. A disadvantage is that the coordinates contained in the catalogue may not correspond to the central star coordinates even for PNe with known central stars. Information about CSPNe is scattered across the literature and often the coordinates themselves are not identified, even for CSPNe that have been studied \citep{weidmannCSPNe}.

In addition to astrometric parameters (positions, parallaxes and proper motions), the \emph{Gaia} satellite measures fluxes in three bands: the wider \emph{G} band covering visible wavelengths and extending partway into the near infrared (330 to 1050nm), and narrower bands \emph{G}$_{BP}$ and \emph{G}$_{RP}$, covering the blue and red halves of the spectrum respectively \citep{gaiaphotometry}. For well-behaved sources the \emph{BP} (blue photometer) and \emph{RP} (red photometer) fluxes essentially add to produce the total flux in \emph{G}, with this degeneracy giving a single measurement of colour as the magnitude difference across any two passbands (e.g. BP -- RP). 

The manner in which the \emph{BP} and \emph{RP} fluxes are measured makes them more susceptible to contamination by nearby sources or errors in background estimation, particularly in densely populated or nebulous regions. Departures from the expected relation between these fluxes and the flux in \emph{G} are indicated by the flux excess factor published in the \emph{Gaia} catalogue; in extreme cases photometry is published for the \emph{G} band only.

The second \emph{Gaia} data release, DR2, contains around 1.7 billion sources, with about 1.3 billion sources having full sets of astrometric parameters (as opposed to positions only) and a similar number (1.4 billion) having full photometry from which colour information can be derived (rather than magnitude in \emph{G} only).

Not all CSPNe will appear as sources in \emph{Gaia} DR2, for a variety of possible reasons: appearing too faint (CSPNe can have high bolometric luminosities but emit most of their light in the ultraviolet), having insufficiently many detections, or being obscured by foreground stars, interstellar dust, or the nebula itself. Indeed the fraction of PNe with secure CSPN identifications is small: \citet{parkercspnpercentage} noted it as 25\%.

CSPNe that are detected may not be the closest sources to the centre of the visible nebula, especially if the full extent of the nebula is not apparent, or in high density regions such as the galactic centre and plane. PN progenitors are hot (blue) stars, but they may not appear blue in the \emph{Gaia} BP -- RP colour space, due to reddening effects or, in the case of binary CSPNe, the presence of a main sequence or giant companion whose light dominates. With binary systems we are still interested in the detected companion as it provides an equally useful parallax measurement. Some bright and compact PNe may themselves appear like stellar sources, though their spectra will be dominated by nebular emission rather than stellar continuum. Again such sources are useful to detect, and we attempt to match them, though they are not, strictly speaking, CSPNe.

For many if not most PNe, our expectation is that the true central star, if it is visible, will be closest to the centre of the nebula. However, many CSPNe will not be detected by \emph{Gaia}, and in those cases the closest stars to the nebula centre will be field stars. Our goal is as much to avoid these impostors as it is to recover true CSPNe, as their inclusion skews any further analysis based on their properties. Thus, a matching approach is required that considers more than just taking the nearest neighbour in the \emph{Gaia} DR2 catalogue for each PN in HASH.

\subsection{Catalogue matching}

We treat the search for CSPNe as a catalogue matching problem, one of finding correspondences between known PNe and sources in \emph{Gaia} DR2. The problem of catalogue matching arises often in astronomy, usually in the context of matching objects detected at different wavelengths. It has been well studied, and has a common solution, the likelihood ratio method (\citet{sutherlandsaunders1992}, henceforth SS92), which provides a principled statistical approach towards determining the reliability of candidate matches. We briefly describe the method here for reference, following SS92.

We suppose that we have a sparse primary catalogue and a dense secondary catalogue, and wish to match objects between them. Given a pair of candidate counterparts in two different catalogues, the idea of the likelihood ratio method is to compare two competing hypotheses: the objects are actually the same (a genuine match), or merely coincidental. In the simplest version, if the positions in the catalogues are offset from each other by an angular separation $r$, the likelihood ratio is the ratio of the probability of finding true counterparts with measured positions separated by $r$ to the probability of finding chance objects with that separation. That is, the likelihood ratio is a ratio of two probability densities. If we assume that Gaussian positional uncertainties are present only in the second catalogue, with standard deviation $\sigma$, the distribution of separations follows a Rayleigh distribution with parameter $\sigma$. Likewise, assuming a constant background density $\rho$, the density of spurious objects at a radial separation $r$ from the primary source position simply increases linearly with $r$ (considering a narrow ring of increasing radius). This gives a likelihood ratio\footnote{The version of the likelihood ratio in Eq. \ref{eq:simpleLR} is related to the figure of merit used in the \emph{Gaia} cross-matches with external catalogues \citep{gaiacrossmatching2017, gaiacrossmatching2019}.}
\begin{equation} \label{eq:simpleLR}
    L = \frac{\textup{Rayleigh}(r; \sigma)}{2\pi r \rho}.
\end{equation}
The likelihood ratio can also incorporate additional properties such as colour and magnitude as well as the prior probability of finding a match. If we consider the colour $c$ in addition to separation $r$, and furthermore assume that they are independent, we get
\begin{equation} \label{eq:LR}
    L = \overbrace{Q}^{\text{prior}} \underbrace{\frac{P(c|\textup{genuine})}{P(c|\textup{chance})}}_{\text{colour term}} \overbrace{\frac{P(r|\textup{genuine})}{P(r|\textup{chance})}}^{\text{separation term}}
\end{equation}
\noindent where $Q$ is the prior probability of there being a match for the object in the secondary catalogue.

While the likelihood ratio is valid for individual sources in isolation, we should consider the likelihood ratios for all candidate matches together for given primary catalogue object. This is done through the reliability, which for the $i$th candidate is
\begin{equation} \label{eq:reliablity}
    R_i = \frac{L_i}{\sum_j L_j + (1 - Q)}.
\end{equation}
Reliability serves as the probability of a match being correct, with the nice properties that the sum of reliabilities of all candidate matches for a given object is at most 1, and the expectation of that sum is the identification rate $Q$.

\subsection{Likelihood ratio method for CSPNe}

In our approach we take HASH as the primary, or leading, catalogue, and the far denser \emph{Gaia} DR2 catalogue as the secondary catalogue. We consider the BP -- RP colour of \emph{Gaia} sources in addition to their positional offsets, motivated by the expectation of PN progenitor stars being hotter and thus having identifiably bluer colours.

The main obstacle to applying the likelihood ratio method is the lack of well-characterised positional uncertainties. Determining priors for colour or other parameters in general does not require a set of verified counterparts, and can be done empirically, using methods such as those in SS92 and \citet{nwaycrossmatching}, which we describe later.

\begin{figure}
    \centering
    \includegraphics[width=\hsize]{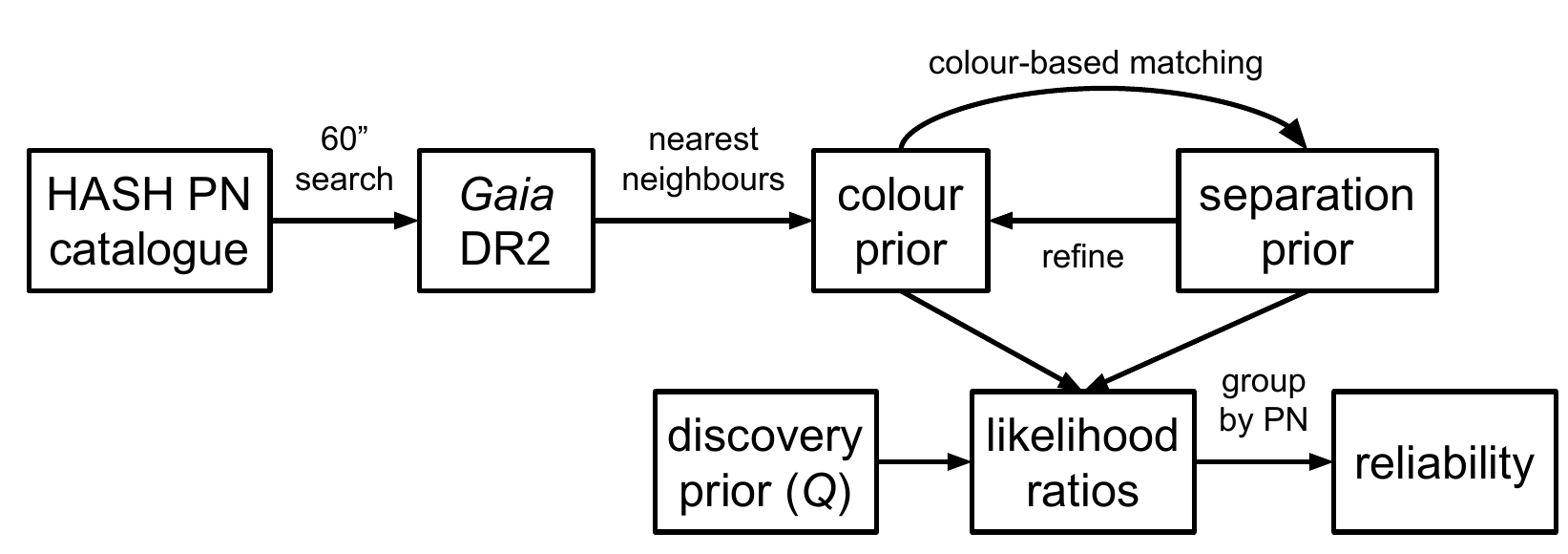}
    \caption{Outline of the steps used in the matching process.}
    \label{fig:flowchart}
\end{figure}

Our approach is iterative, involving first determining an approximate colour prior based on a simple positional cross-match. Using this prior, we select a new set of counterparts that have high confidence based on colour alone. The angular separations of these counterparts are used to determine positional uncertainties, which in turn are used to generate an improved colour prior to be used in the final matching (Fig. \ref{fig:flowchart}).

Though not a strict application of the method, this approach borrows some of the ideas of the co-training technique used in semi-supervised learning \citep{cotraining}, in which classifiers based on different views of a set of data label examples in order to train each other. It can also be seen as considering the colour and separation terms in Eq. \ref{eq:LR} separately and alternating between them.

\subsubsection{Nearest neighbour selection}

\begin{figure}
    \centering
    \includegraphics[width=\hsize]{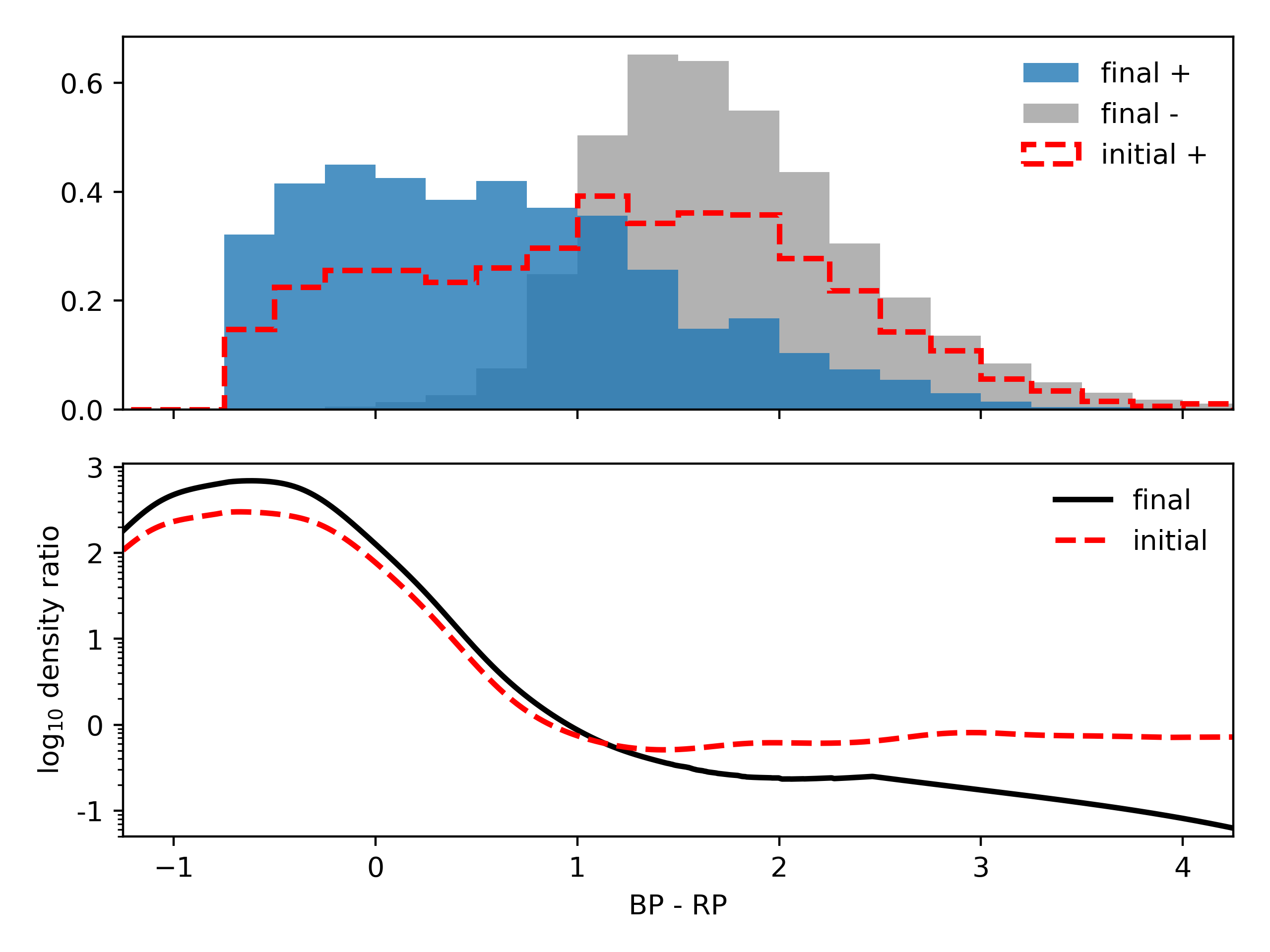}
    \caption{Histogram of BP -- RP colours for three different sets of \emph{Gaia} sources covering both iterations of the colour density ratio computation. The grey histogram shows the distribution of colours of background sources, which does not change visibly between iterations. The initial colour distribution derived from nearest neighbouring sources is indicated by the red dashed line.
    The colour distribution of the final selection based on separation is indicated in blue. The lower panel shows the density ratio in colour space, also for both iterations, with the final density ratio indicated by the black line, and the initial ratio derived nearest versus non-nearest neighbours shown by the dashed red line. All densities are for sources with a well-behaved BP/RP excess factor (indicating reliable colour measurements).}
    \label{fig:colourhistogram}
\end{figure}

We select all \emph{Gaia} sources within 60\arcsec of the roughly 2500 confirmed PNe in HASH (\texttt{PNstat = T}). We take the closest \emph{Gaia} source to each PN location (applying a generous separation cutoff of half the PN radius plus 2\arcsec), and compare the empirical distribution of BP -- RP colours of these nearest neighbour sources with the other (non-nearest neighbour) sources (sources with no BP -- RP colour are ignored in this selection). A simplified version of this comparison is shown in Fig. \ref{fig:colourhistogram}.

In practice we use kernel density estimation rather than the histograms directly, in order to produce a smooth density ratio function, and also consider the BP/RP excess factor as in indicator of the uncertainty in the colour measurement. More details can be found in Appendix \ref{impdet-cdrest}.

While the caveats we mentioned previously apply, we do expect that most true central stars will be the nearest sources to the PN centres, and that a significant fraction of nearest neighbours will indeed be true central stars. Thus the nearest neighbour and non-nearest neighbour colour distributions should approximate $P(c|\textup{genuine})$ and $P(c|\textup{chance})$, with some contamination in both directions. The effect of the contamination is to push the ratio of these densities towards unity, but the structure should still be preserved.

S92 suggested a similar approach based on taking all possible counterparts within a $3\sigma$ positional error ellipse, and subtracting from the derived density a representative sample of background objects to account for the expected contamination. The equivalent here would be to subtract the non-nearest neighbour colour distribution from the nearest neighbour one, though we deem this unnecessary for our application.

We use this initial colour prior to select the subset of \emph{Gaia} sources within our radius cutoff (not necessarily nearest neighbours) that are high confidence matches based on colour alone. When there are multiple sources with strongly suggestive colours for a single PN, we take the source with the smallest angular separation.

We assume that, for true counterparts, the apparent colour of the CSPN (which may be affected by extinction or binarity) and its position relative to the nominal centre of its PN are independent of each other.\footnote{This assumption may not always hold true in practice, especially if blue CSPNe are preferentially used as PN positions in catalogues. Also, colour and separation will certainly not be independent in the final catalogue, as both features are used in the selection.} Under this assumption, the positional distribution of these colour-selected counterparts should be representative of the positional distribution of all true counterparts, and we can use the former as an estimate of the latter. This is what we do in the next step.

\subsubsection{Positional uncertainty and background density estimation} \label{positionaluncertainty}

\begin{figure}
    \centering
    \includegraphics[width=\hsize]{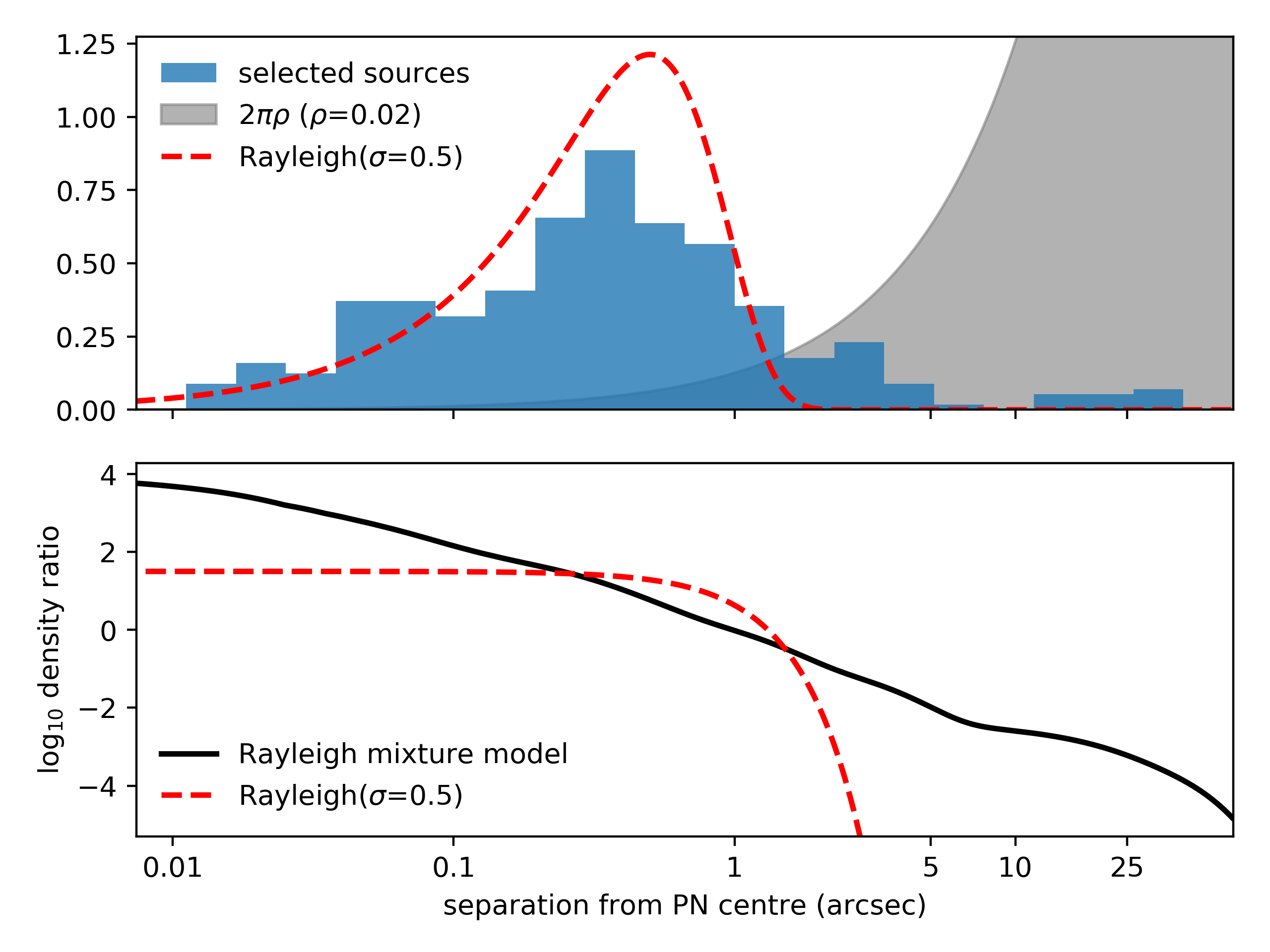}
    \caption{Histogram of the separations of the "high-confidence" sources (selected by colour) from their PN centres, along with, for comparison, a Rayleigh distribution with a similar mode in red, and a uniform density of background sources in grey. The lower panel shows separation density ratio resulting from the derived mixture of Rayleigh distributions compared to that from the single Rayleigh distribution in the upper panel. In practice the mixture is re-weighted depending on the radius of the PN.}
    \label{fig:separationhistogram}
\end{figure}

The angular separations from the sources selected by colour to their PN centres (upper half of Fig. \ref{fig:separationhistogram}) range from fractions of an arcsecond to tens of arcseconds, and the concentration of very nearby sources combined with the long tail is not well described by a single Rayleigh distribution as in Eq. \ref{eq:simpleLR}. This is not unexpected as there are many possible sources of disagreement between the PN position and that of its central star:

\begin{itemize}
    \item catalogue inhomogeneity, in particular whether the catalogue position was based on nebula or central star
    \item for positions based on central star positions, inherent uncertainty in that measurement
    \item for positions not based on central stars, uncertainties as to the true location of the nebula centre, and also the possibility of offset due to relative motion between the central star and the nebula
    \item effects of proper motion (different measurement times)
    \item \emph{Gaia} measurements errors (negligible relative to to other sources of uncertainty, so we do not include the \emph{Gaia} errors explicitly)
\end{itemize}

Thus to estimate the distribution of PN centre separations for true CSPNe ($P(r|\textup{genuine})$), we fit a mixture of Rayleigh distributions, with one distribution per PN -- \emph{Gaia} source pair in our colour selection. Each individual distribution is fit to the maximum likelihood parameter for the angular separation between the \emph{Gaia} source and the PN centre. This construction approach simplifies the estimate and ensures the the separation density ratio is smooth, strictly decreasing, and behaves well near zero.

Some sources of uncertainty are dependent on the nebula size. Thus we re-weight the mixture to reflect offsets to CSPNe for PNe of similar sizes. For example, for a PN with a radius of 60\arcsec the mixture will be be dominated by PN -- \emph{Gaia} source pairs where the radius of the PN is between around 30\arcsec and 120\arcsec.

The other component of the term in the angular separation likelihood ratio is the density of background sources, $\rho$. We estimate this locally for each PN by counting the \emph{Gaia} sources found within the 60\arcsec search window. We choose this approach over other methods (e.g. taking the separation to the $n$th nearest object) for its simplicity. The background source density will be the same for all candidate CSPNe for a given PN, so it does not affect the relative ranking, only the confidence.

\subsubsection{Colour prior refinement}

The derived positional uncertainties can in turn be used to generate a new colour prior, in a manner similar to the "self-constructed priors" described in \citet{nwaycrossmatching}. The idea is to derive priors based on the properties of counterparts who identification is reasonably secure based on position alone. Instead of nearest and non-nearest neighbours, we now use the colours of positionally selected CSPNe and non-CSPNe to determine $P(c|\textup{genuine})$ and $P(c|\textup{chance})$ respectively, leaving out sources for which the position by itself is inconclusive.\footnote{We calculate reliability based on separation only, using sources with reliability > 0.8 as our positive examples and those with reliability < 0.2 as negative ones.} This removes many of the contaminants from the previous estimation based on nearest neighbour, showing a stronger preference for blue colours and a decreased score assigned to redder \emph{Gaia} sources -- in essence, increasing the contrast in the colour density ratio function.

In principle we could alternate back and forth between updating the distances and colour distributions, but the updated colour prior does not significantly change which sources meet the threshold used for the selection at the end of Sect. \ref{positionaluncertainty}. Thus further iteration is not necessary.

\subsubsection{Final steps}

The final piece of the likelihood ratio function is $Q$, the identification rate. This scales all likelihood ratios, but does not change the ranking. We choose a value for $Q$ of 0.5, which we will verify later.

We calculate the likelihood ratios for all \emph{Gaia} sources within each 60\arcsec search window, though in selecting matches we enforce an additional separation cutoff of half the radius of the PN plus 2\arcsec. We do this for all confirmed, likely, and possible PNe in HASH, though only the confirmed PNe were used in deriving the priors.

Sources missing BP--RP colours in \emph{Gaia} have likelihood ratios computed based on position only (equivalent to having the colour term in the likelihood ratio being equal to 1).

Following SS92, we compute the reliability of candidate sources for each PN, using that as our scoring metric.

\section{Matching results} \label{results}

\begin{figure}
    \centering
    \includegraphics[width=\hsize]{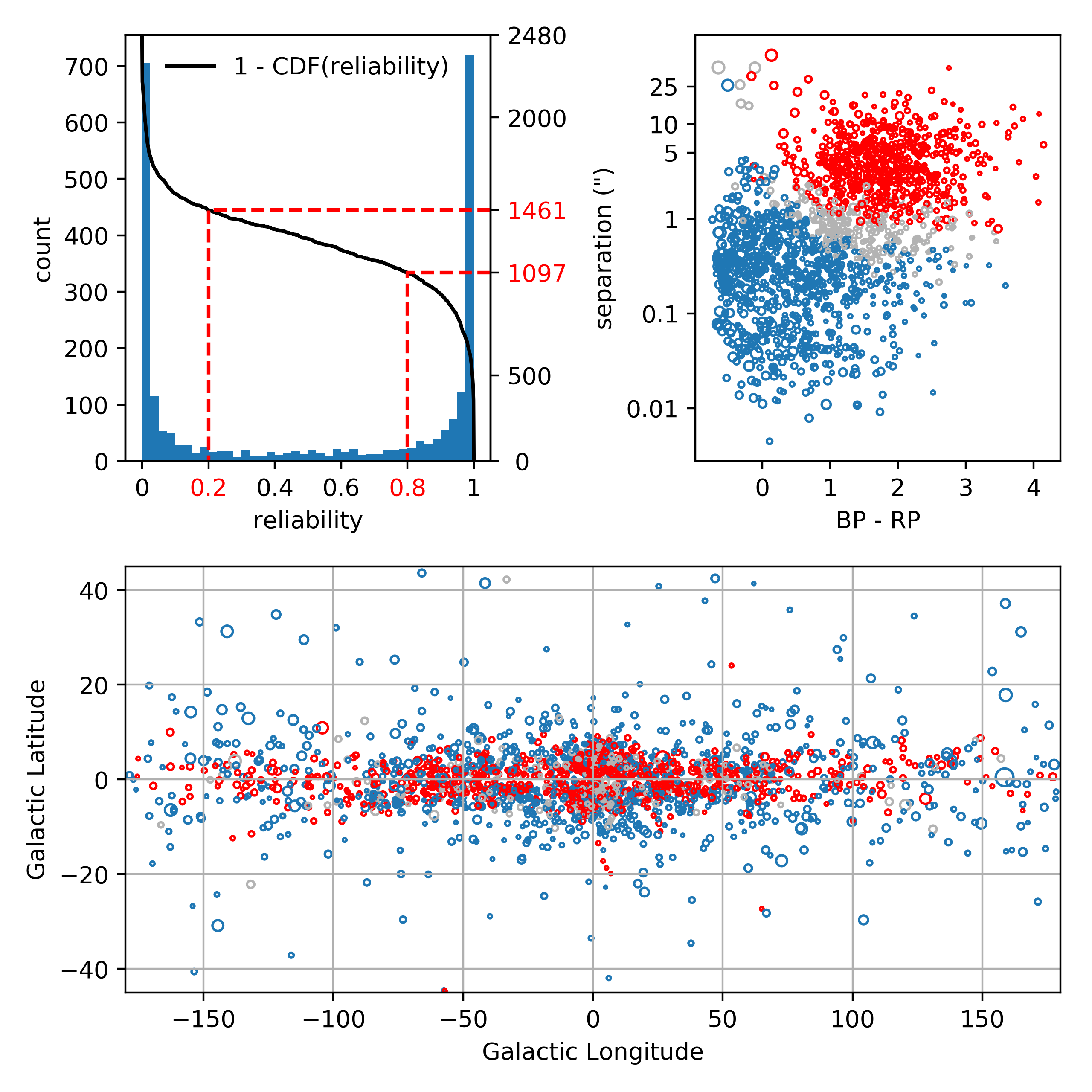}
    \caption{Matching results for confirmed PNe in HASH. The histogram on the upper left shows the reliabilities of highest ranked candidate central stars for each PN. Overplotted is the mirrored cumulative distribution function (CDF) of that distribution, with the cutoffs and counts for best and potential matches highlighted. The two scatter plots show the distribution of the matches in colour / separation space and in galactic coordinates, with blue circles being likely matches, grey circles being possible matches, and red circles being rejected sources. Larger circles correspond to PN with larger angular sizes.}
    \label{fig:results}
\end{figure}

The reliability distribution of the highest ranked candidate for each PN is strongly bimodal (Fig. \ref{fig:results}, upper left), meaning that for most PNe our method has either selected a single \emph{Gaia} source as the best central star candidate with high confidence or rejected all nearby sources. The mean sum of reliabilities for the 2480 confirmed PNe in HASH is 0.53, consistent with our chosen value of $Q$.\footnote{Otherwise we could iteratively update $Q$ and recalculate the reliabilities until they converge. Indeed, the 415 likely and 663 possible PNe have mean reliabilities of 0.45 and 0.34 respectively, indicating a lower success rates for these unconfirmed PNe and also that the chosen $Q$ value of 0.5 is thus inconsistent for them. The relationship between PN status and central star matching success is expected given that a clearly visible central star contributes towards confirming a nebula's PN status.} We focus the remainder of our analysis on these confirmed PNe, as they are most relevant for scientific applications.

Based on the shape of the reliability distribution, we choose 0.8 as our threshold for likely matches and 0.2 as our threshold for possible matches. Applying these thresholds, we find 1086 likely matches and 381 possible matches, representing 44\% and 15\% respectively of the total number of confirmed PNe.

The highest confidence matches are \emph{Gaia} sources that are both blue and within fractions of an arcsecond of the HASH position. However either of these criteria alone can be sufficient - our method also finds more distant blue sources and accepts red sources that are very central (Fig. \ref{fig:results}, upper right).

The greatest matching success rates are for extended PNe away from the galactic centre and away from the disc (lower half of Fig. \ref{fig:results}), where the PNe tend to be nearer, the density of background objects lower, and the visible light from stars less reddened by dust. Most of the uncertain matches are towards the galactic centre; cursory inspection of these shows that many are missing colours in \emph{Gaia} and that their positional offsets are too large to accept the candidates based on position alone. These could benefit from the incorporation of additional photometry from other surveys such as VPHAS+ \citep{vphassurvey}, spectroscopic followup, or, for those that do have \emph{Gaia} colours, reddening estimates. It is interesting to observe that few PNe have multiple plausible candidate CSPNe; the choice is generally between a single best candidate and the conclusion that the CSPN as not been detected by \emph{Gaia} at all.
 
\subsection{Comparison with previous works} \label{comparison}

It is difficult to find large samples of positions of verified CSPNe in the literature that would serve as a ground truth to which to compare our matching results. \citet{kerber03} (henceforth K03) listed 201 PNe positions based on central stars, but those identifications were based on visual assessment of single-band broadband images, and the coordinates had potentially large uncertainties. The more recent catalogue of \citet{weidmannCSPNe} collected 492 PNe that had CSPN spectral types in literature; however the authors themselves noted that the positions in their catalogue were generally those of the PNe rather than of the CSPNe, which gives them limited utility for cross-matching.\footnote{We do note that our method failed to find high reliability matches for 20\% of the 412 PNe in the \citet{weidmannCSPNe} catalogue listed as confirmed PNe in HASH, which provides some idea of the completeness of \emph{Gaia} DR2.}

The most straightforward comparisons to make are to previously published catalogues based on \emph{Gaia} DR2 because the \emph{Gaia} source ID is unambiguous. That is the focus of this section. However it is important to note that those catalogues are not themselves sources of verified CSPNe, and indeed it is our hope that the matching procedures we developed in this work lead to higher accuracy in many cases.

\subsubsection{Methods}

We compare our matching results to the published catalogues from \citet{kimeswenger2018} (henceforth KB18), \citet{gapn} (henceforth GS19), and \citet{stanghellini2020} (henceforth S+20). KB18 performed manual cross-matching of PNe with radio distances from \cite{sh10} (henceforth SH10), based on literature sources and imagery, while the latter two works relied on purely positional cross-matches based on a variety of literature sources. The matching objectives of these works mirrored ours: to find \emph{Gaia} DR2 sources corresponding to CSPNe (or stellar-like PNe) and to use their parallaxes as distance indicators. We note the overall rates of agreement with our catalogue, and have spot-checked many examples that have good imagery available, but a full comparison is beyond the scope of this work, especially as a large verified CSPN sample is lacking.

KB18 published all 382 of their matches. GS19 claimed to have matched 1571 PNe, but published only a "golden sample" of 211 sources, which had additional quality cuts applied, including a maximum relative parallax error of 30\%. Finally, S+20 limited their published sample to the 430 of the 655 sources they matched that had positive parallaxes with relative errors better than 100\%.

We further filter the published catalogues to those matches that are for confirmed PNe in HASH and whose positions are within our 60\arcsec search radius around the HASH position. Between 50\% and 75\% of the eliminated matches are listed as non-PNe (Symbiotic Stars, HII regions, etc.) in HASH, while the remainder have unconfirmed PN status. Our search radius means that we do miss the central stars of three very large (700" to 1900" diameter) asymmetric PNe in GS19; the SIMBAD positions of these objects correspond to blue central stars, which we believe to be correct, while the HASH positions are significantly offset. These are Sh 2-188 (PN G128.0-04.1), FP J1824-0319 (PN G026.9+04.4), and FP J0905-3033 (PN G255.8+10.9). The same match for Sh 2-188 is also present in S+20, as is a very highly offset match for the compact PN M 1-55 (PN G011.7-06.6). We believe the latter to be due to an unintentional confusion of that PN's coordinates with those of a nearby guide star by K03, which S+20 cite as the source of many of their positions.

\begin{figure}
    \centering
    \includegraphics[width=\hsize]{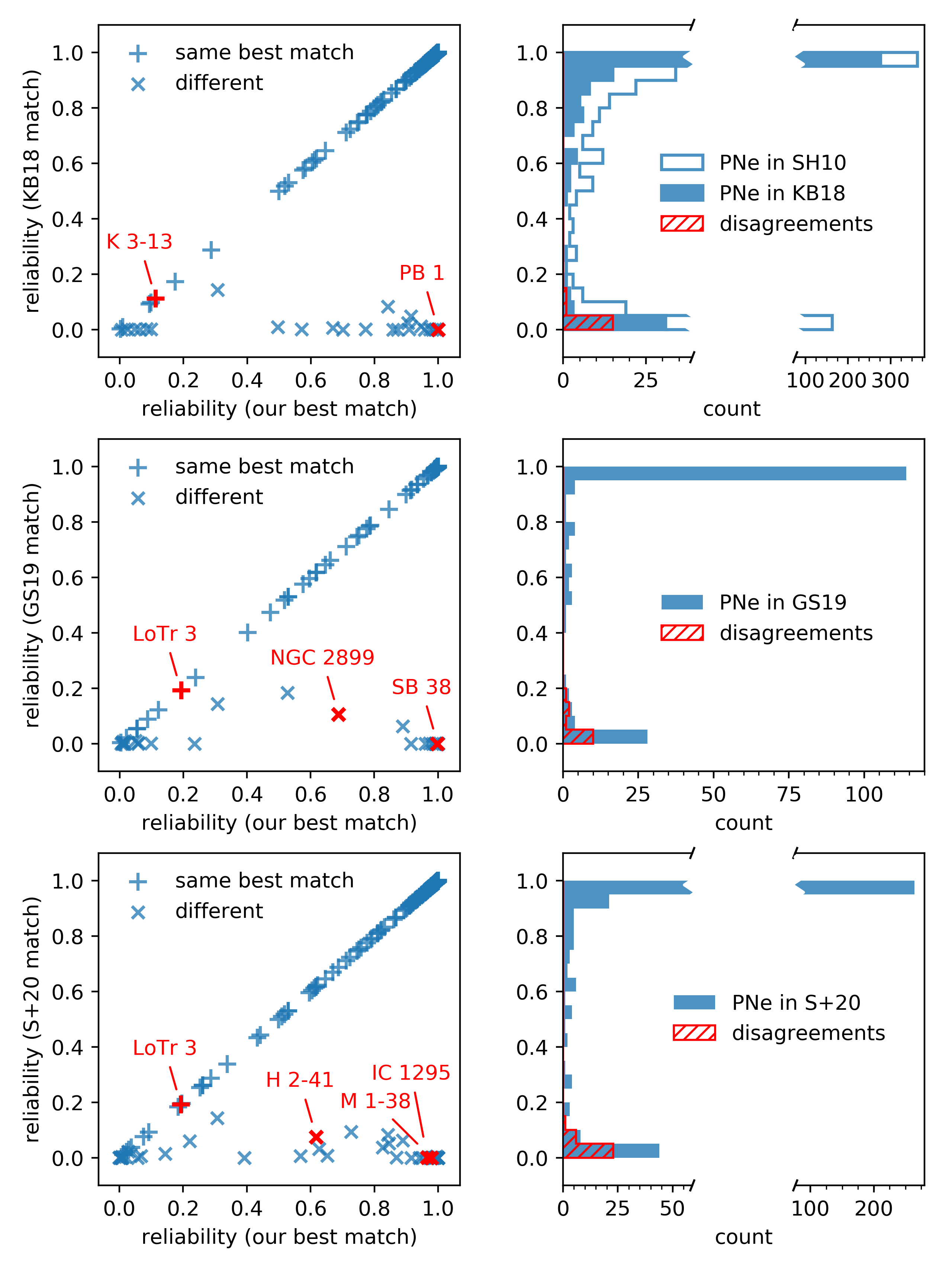}
    \caption{Reliabilities of our best candidate central star matches for confirmed PNe in HASH compared to the reliabilities of the matches for those same PNe published by KB18 (top), S+20 (middle), and GS19 (bottom). The histograms reflect the total counts of reliabilities of matches from previous works, with the top histogram for KB18 also including reliabilities of best matches for all PNe in SH10, including those for which KB18 did not find matches.}
    \label{fig:kb18reliability}
\end{figure}

\begin{table}
\caption{\label{countstable}Comparison counts with previous works.}
\centering
\begin{tabular}{lccccc}
\hline\hline
\smallskip
Reference&Total&\multicolumn{2}{c}{Reliability}&Disagreements\\
&&>0.8&<0.2&\\
\hline
\smallskip
\textbf{KB18}&362&305&37&22\\
&&(84\%)&(10\%)&(7\%)\\
blue sources\tablefootmark{a}&82&80\tablefootmark{b}&0&0\\
&&(98\%)&(0\%)&(0\%)\\
$\sigma_\omega/\omega<0.15$&76\tablefootmark{c}&68&3&2\\
&&(89\%)&(3\%)&(2\%)\\
$\sigma_\omega/\omega<0.25$&106\tablefootmark{c}&89&9&4\\
&&(84\%)&(8\%)&(4\%)\\
\hline
\smallskip
\textbf{GS19}&175\tablefootmark{d}&120&37&18\\
&&(69\%)&(21\%)&(10\%)\\
\hline
\smallskip
\textbf{S+20}&380\tablefootmark{e}&295&56&35\\
&&(78\%)&(15\%)&(9\%)\\
$\sigma_\omega/\omega<0.2$&131&100&19&13\\
&&(76\%)&(15\%)&(10\%)\\
\hline
\end{tabular}
\tablefoot{
Counts consider only confirmed PNe in HASH where the reference \emph{Gaia} source is within 60\arcsec of the HASH position. Disagreements are PNe for which our method found a different, plausible best match (reliability > 0.2), which is independent of the reliability assigned to the reference match (so most disagreements contribute also to the count of reliability < 0.2 matches). The symbol $\omega$ is used for parallax, so $\sigma_\omega/\omega$ is relative parallax error.\\
\tablefoottext{a}{Sources with -0.65 < BP--RP < -0.25.}
\tablefoottext{b}{There are two blue sources with high angular separation that we believe to be correct in KB18; they were selected by our method but with lower reliability on account of their offsets.}
\tablefoottext{c}{Following KB18, these samples are also restricted to parallax derived distances within 4kpc.}
\tablefoottext{d}{Not including three sources outside our 60\arcsec search radius. See text.}
\tablefoottext{e}{Not including two sources outside our 60\arcsec search radius. See text.}
}
\end{table}

\begin{figure*}
    \centering
    \begin{subfigure}[t]{0.33\hsize}
        \centering
        \includegraphics[width=0.45\hsize]{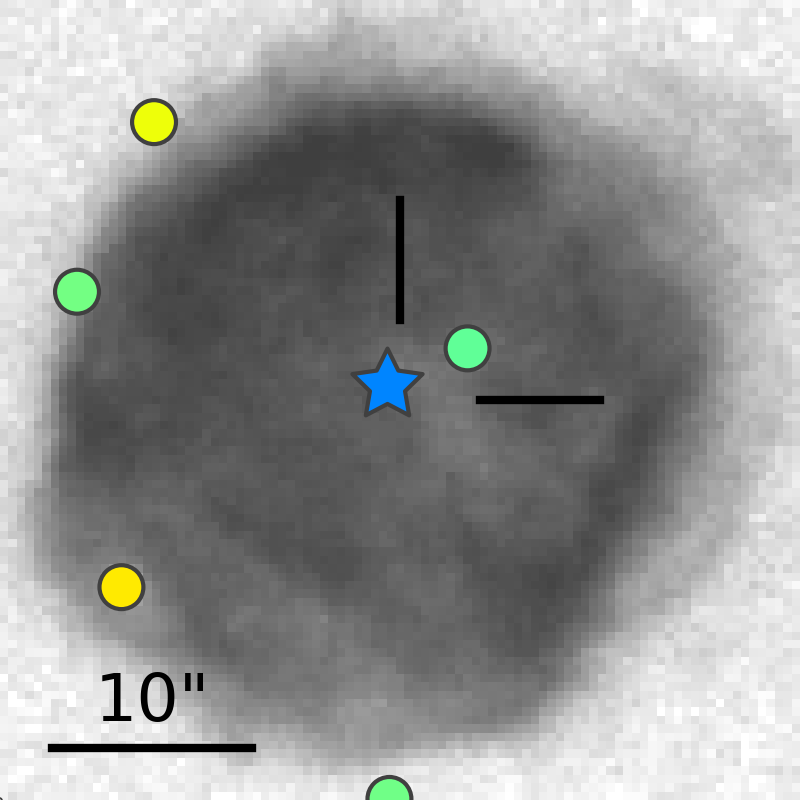}
        \enskip
        \includegraphics[width=0.45\hsize]{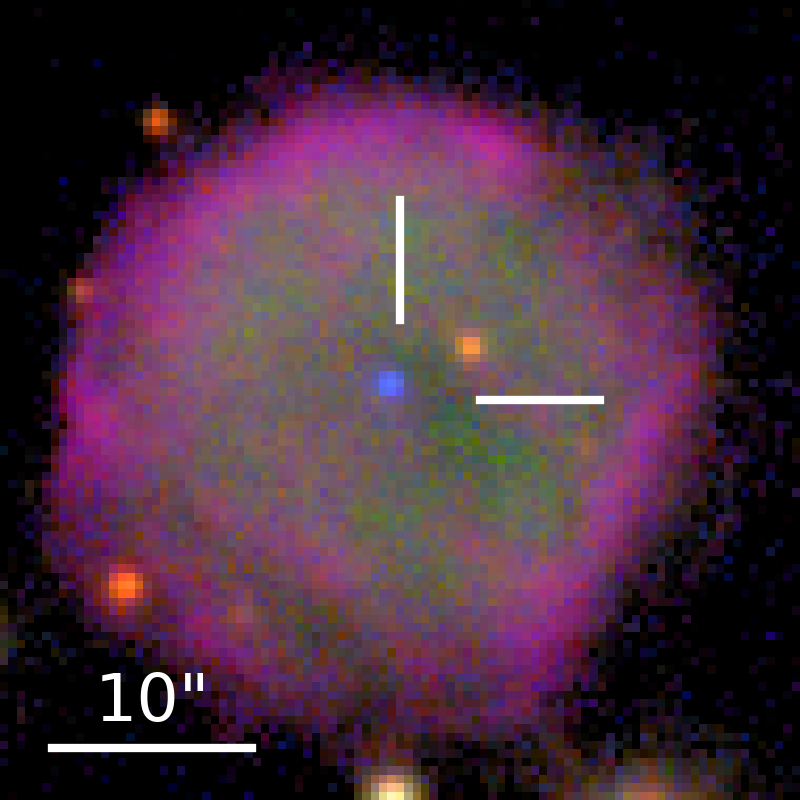}
        \caption{\label{fig:kb18comparison:m118}M 1-18 (PN G231.4+04.3)}
    \end{subfigure}
    \begin{subfigure}[t]{0.33\hsize}
        \centering
        \includegraphics[width=0.45\hsize]{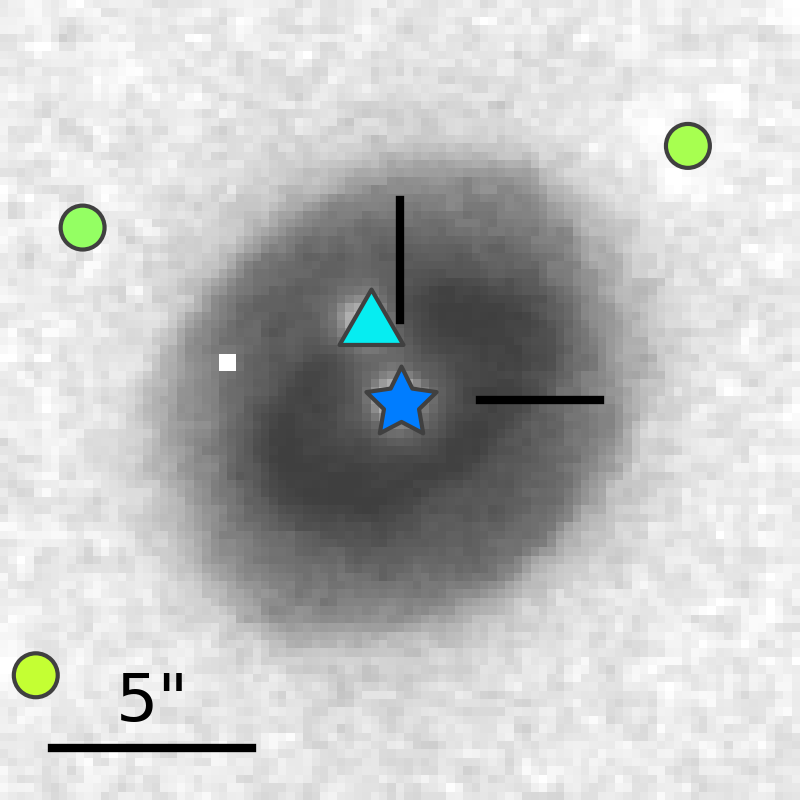}
        \enskip
        \includegraphics[width=0.45\hsize]{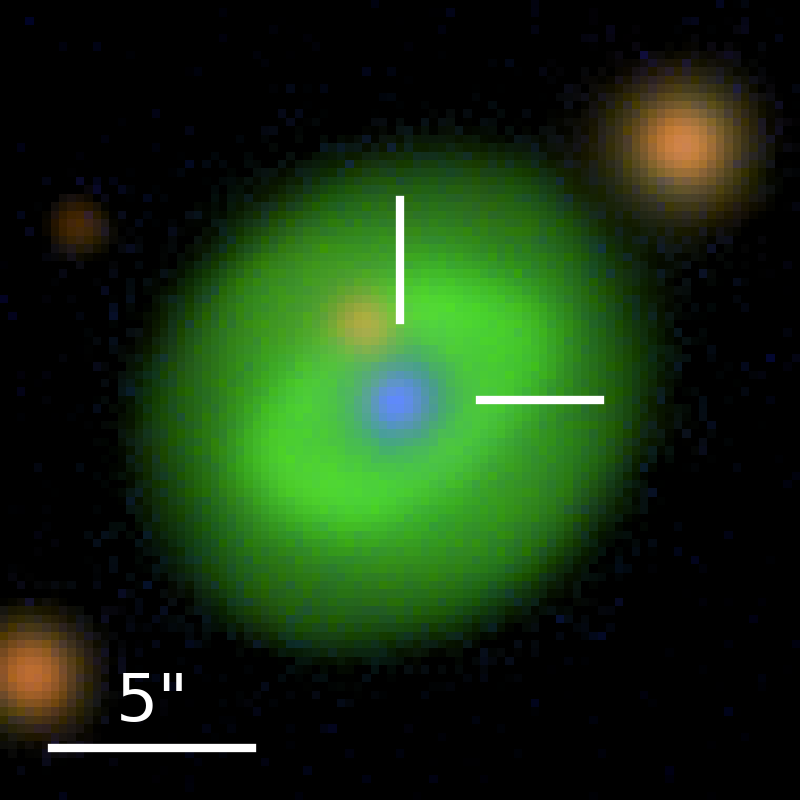}
        \caption{\label{fig:kb18comparison:pb1}PB 1 (PN G226.4-03.7)}
    \end{subfigure}
    \begin{subfigure}[t]{0.33\hsize}
        \centering
        \includegraphics[width=0.45\hsize]{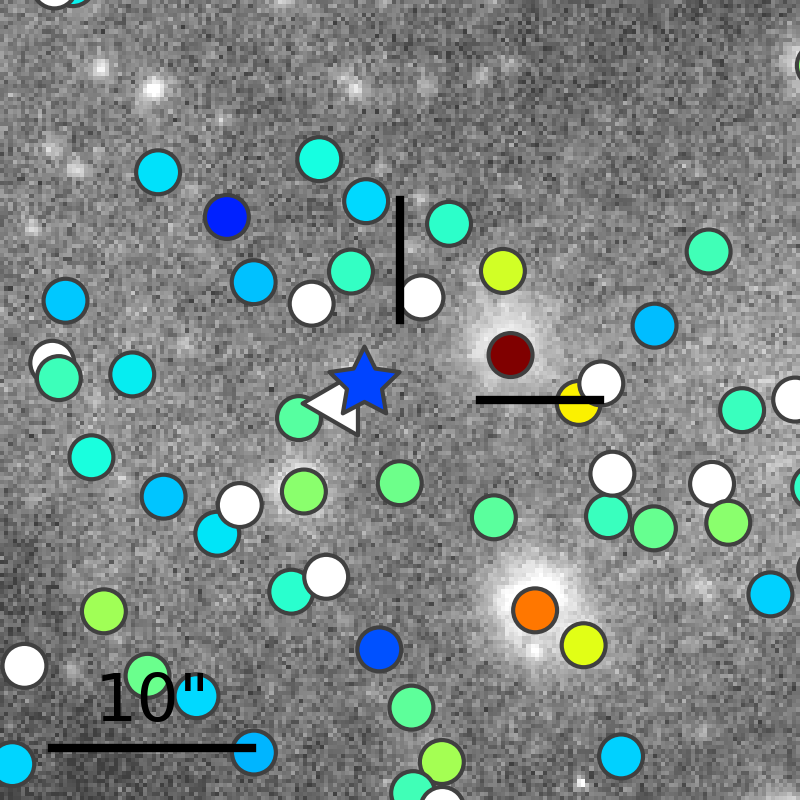}
        \enskip
        \includegraphics[width=0.45\hsize]{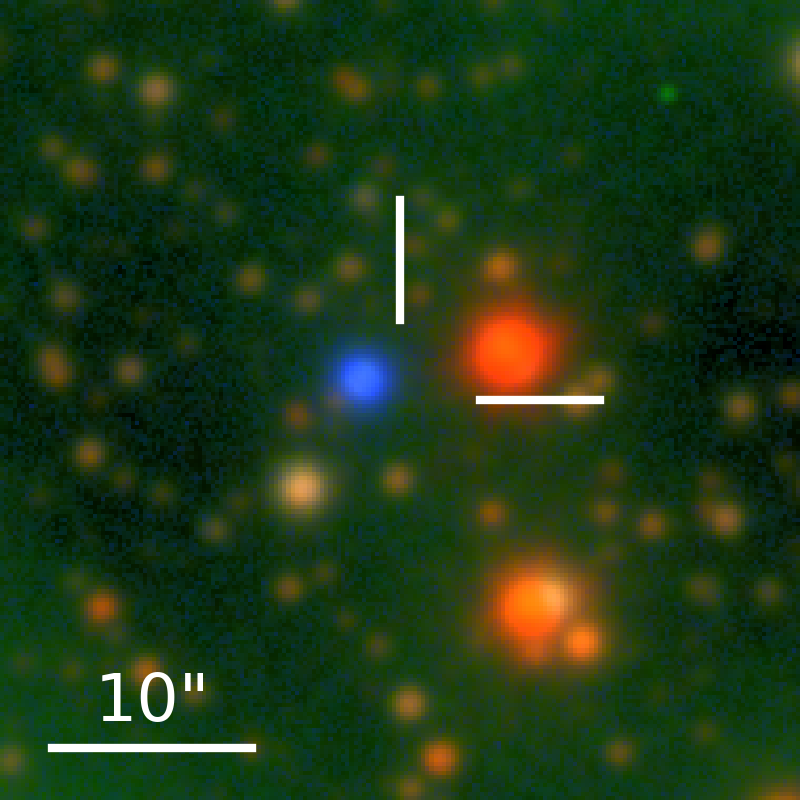}
        \caption{\label{fig:kb18comparison:ic1295}IC 1295 (PN G025.4-04.7)}
    \end{subfigure}
    \par\medskip
    \begin{subfigure}[t]{0.33\hsize}
        \centering
        \includegraphics[width=0.45\hsize]{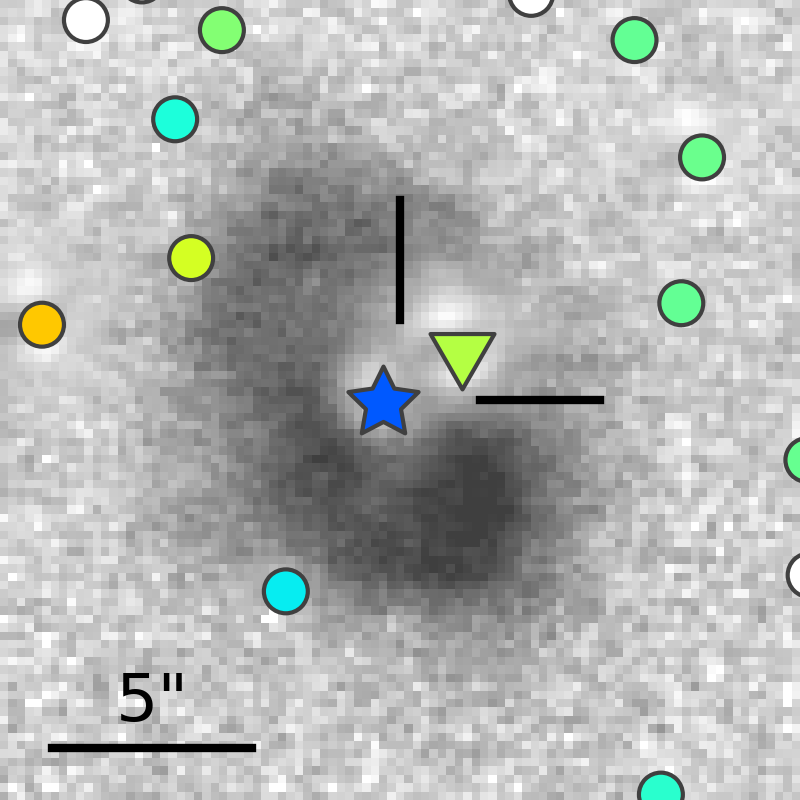}
        \enskip
        \includegraphics[width=0.45\hsize]{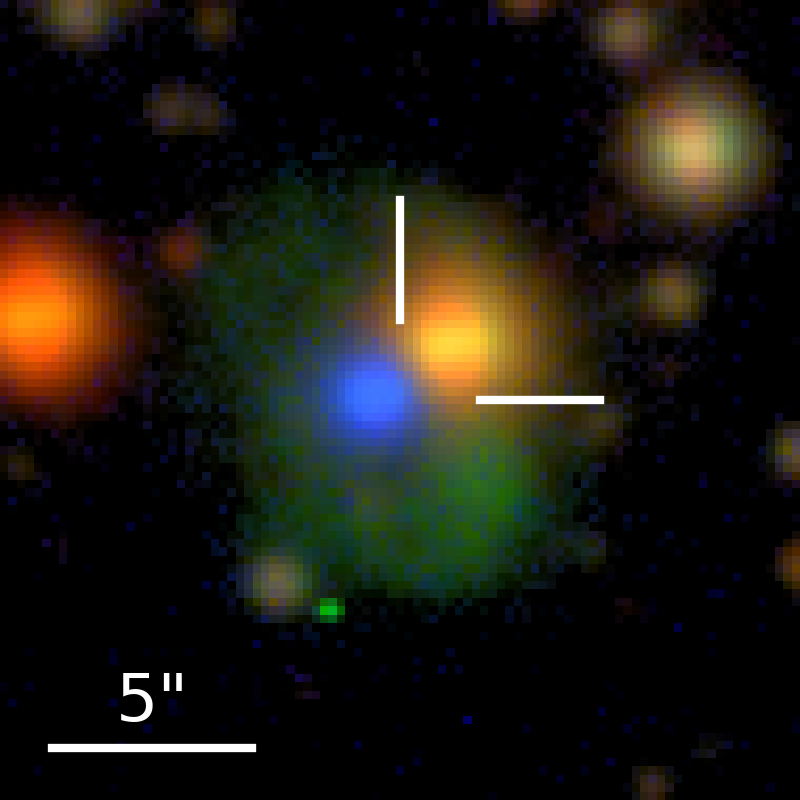}
        \caption{\label{fig:kb18comparison:sb38}SB 38 (PN G352.7-08.4)}
    \end{subfigure}
    \begin{subfigure}[t]{0.33\hsize}
        \centering
        \includegraphics[width=0.45\hsize]{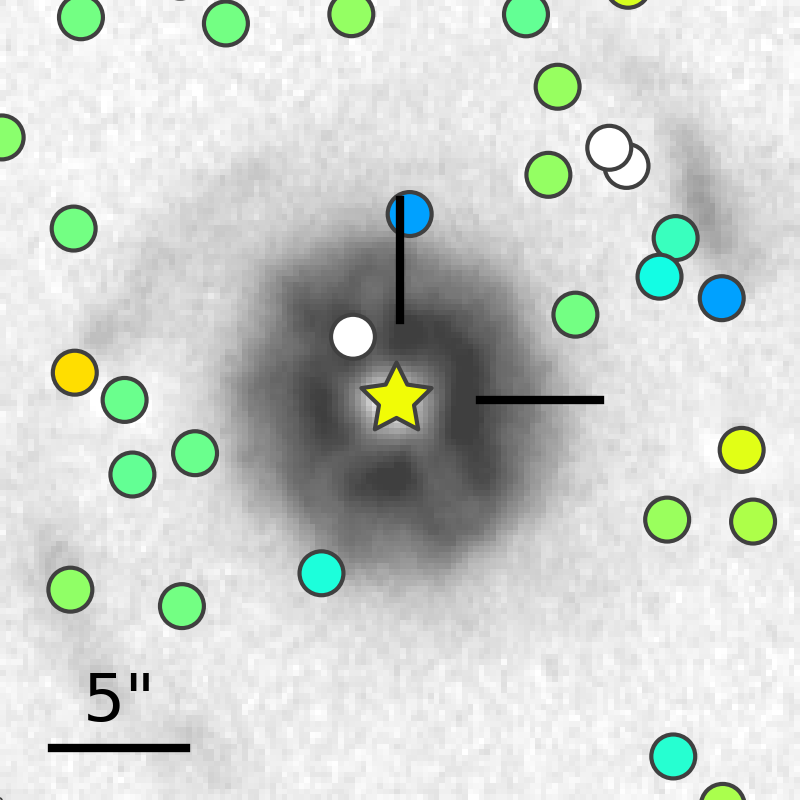}
        \enskip
        \includegraphics[width=0.45\hsize]{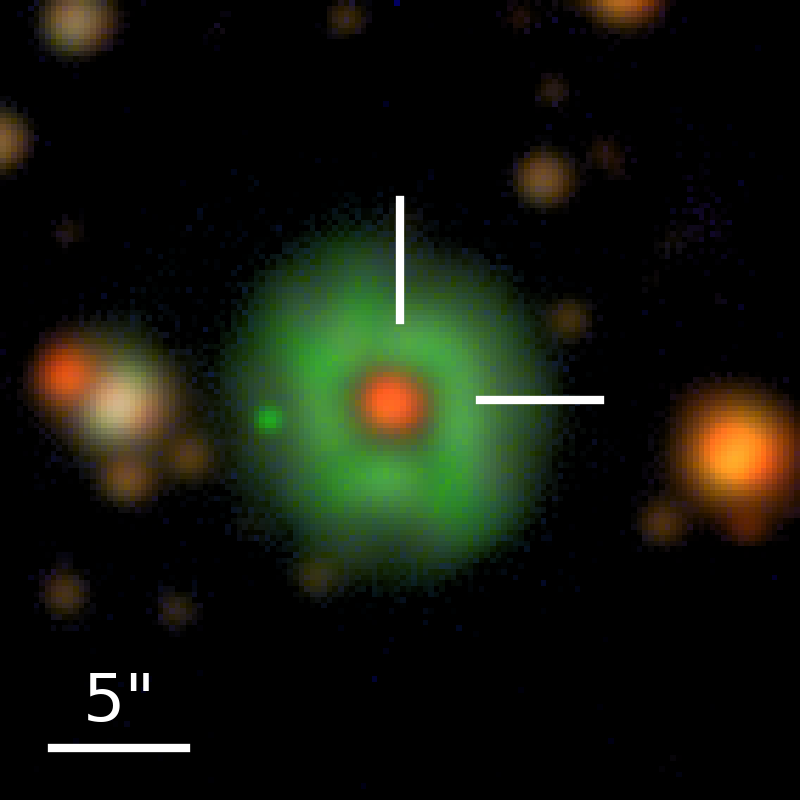}
        \caption{\label{fig:kb18comparison:hen239}Hen 2-39 (PN G283.8-04.2)}
    \end{subfigure}
    \begin{subfigure}[t]{0.33\hsize}
        \centering
        \includegraphics[width=0.45\hsize]{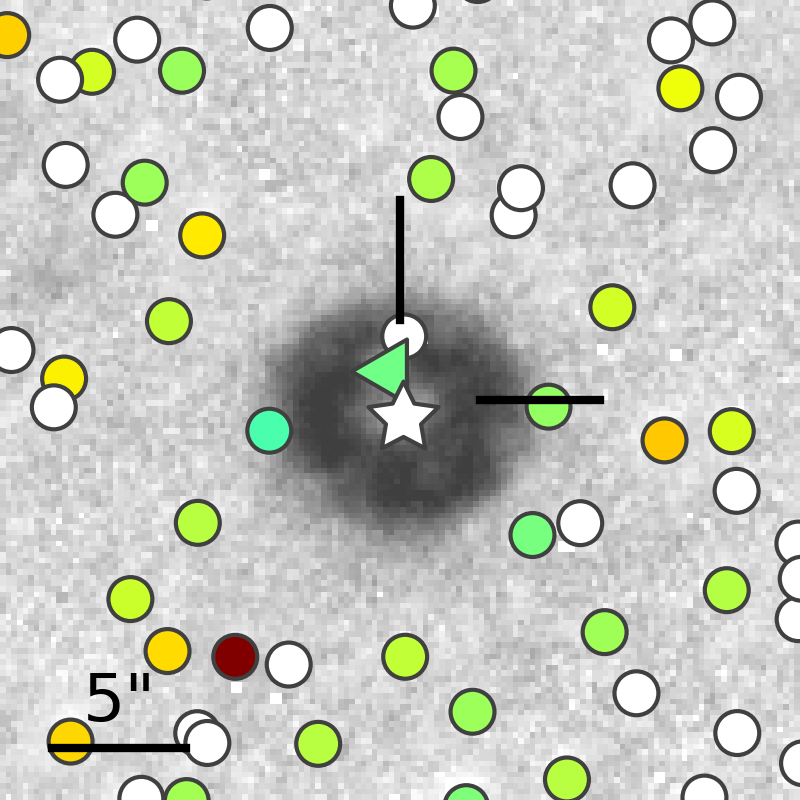}
        \enskip
        \includegraphics[width=0.45\hsize]{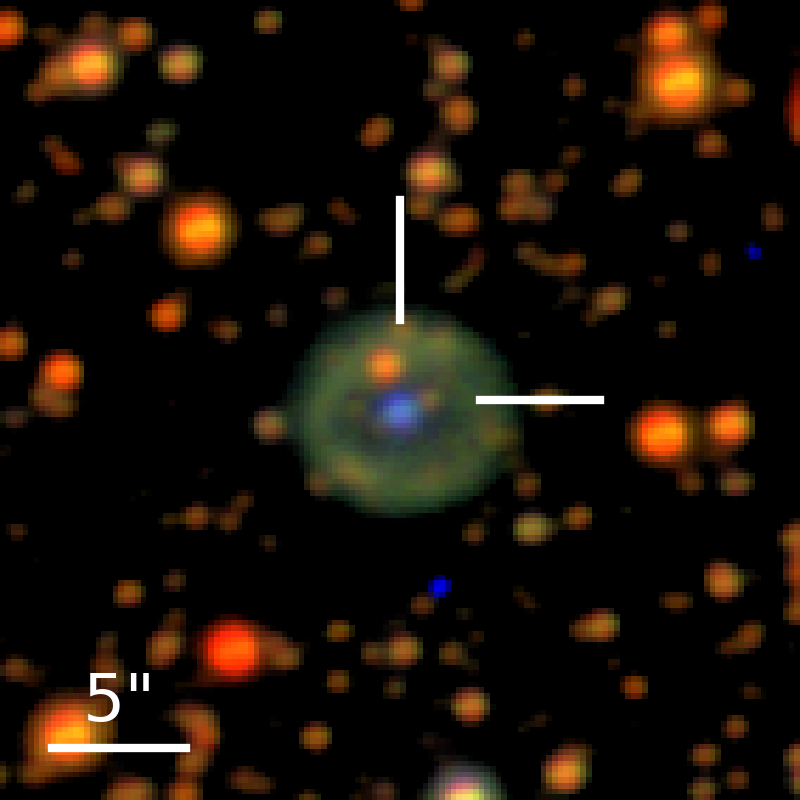}
        \caption{\label{fig:kb18comparison:h241}H 2-41 (PN G003.8-04.5)}
    \end{subfigure}
    \par\medskip
    \begin{subfigure}[t]{0.33\hsize}
        \centering
        \includegraphics[width=0.45\hsize]{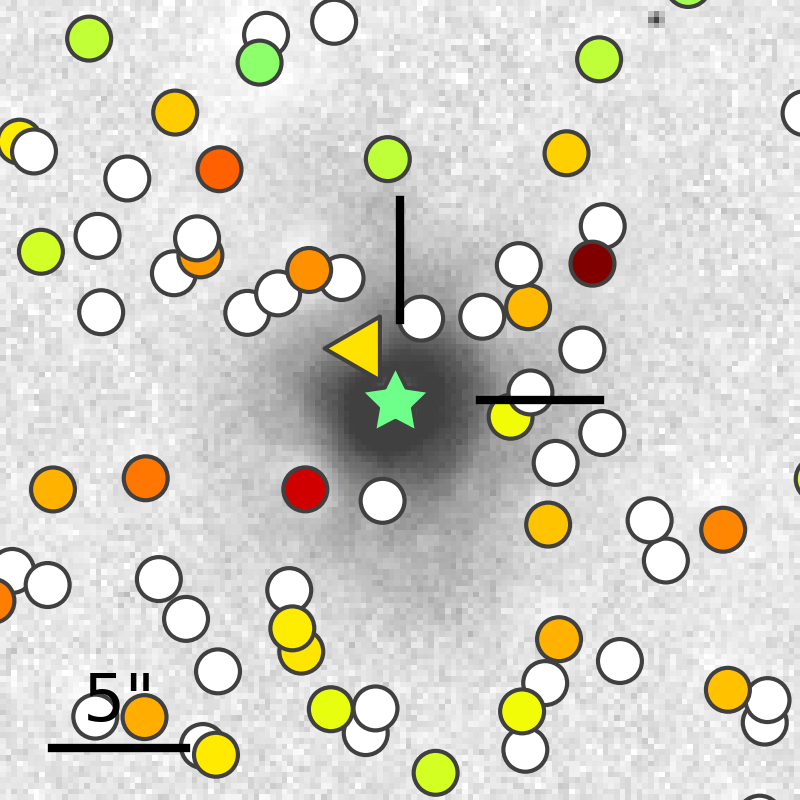}
        \enskip
        \includegraphics[width=0.45\hsize]{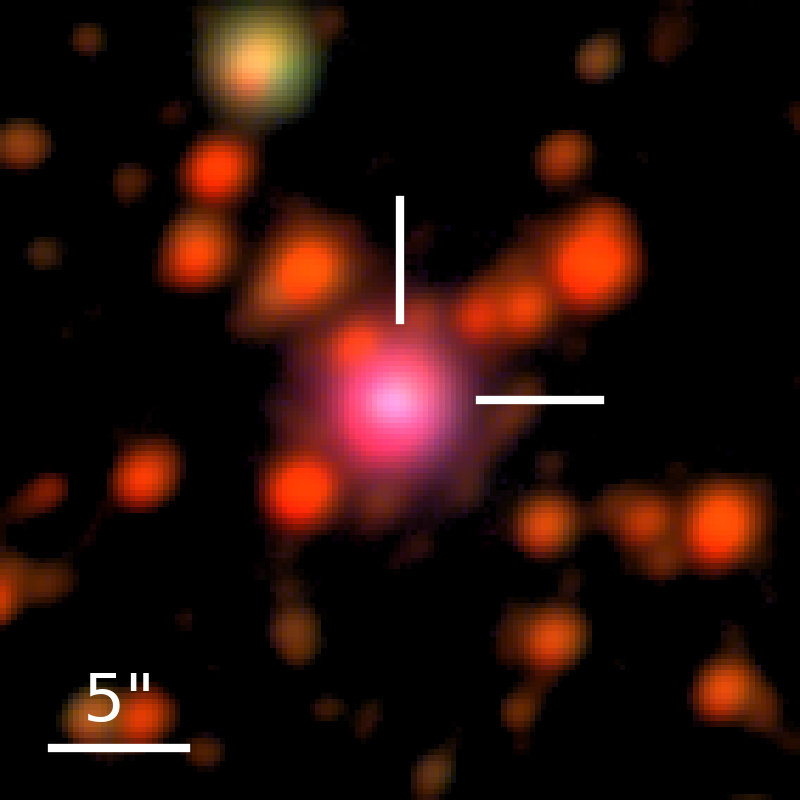}
        \caption{\label{fig:kb18comparison:m138}M 1-38 (PN G002.4-03.7)}
    \end{subfigure}
    \begin{subfigure}[t]{0.33\hsize}
        \centering
        \includegraphics[width=0.45\hsize]{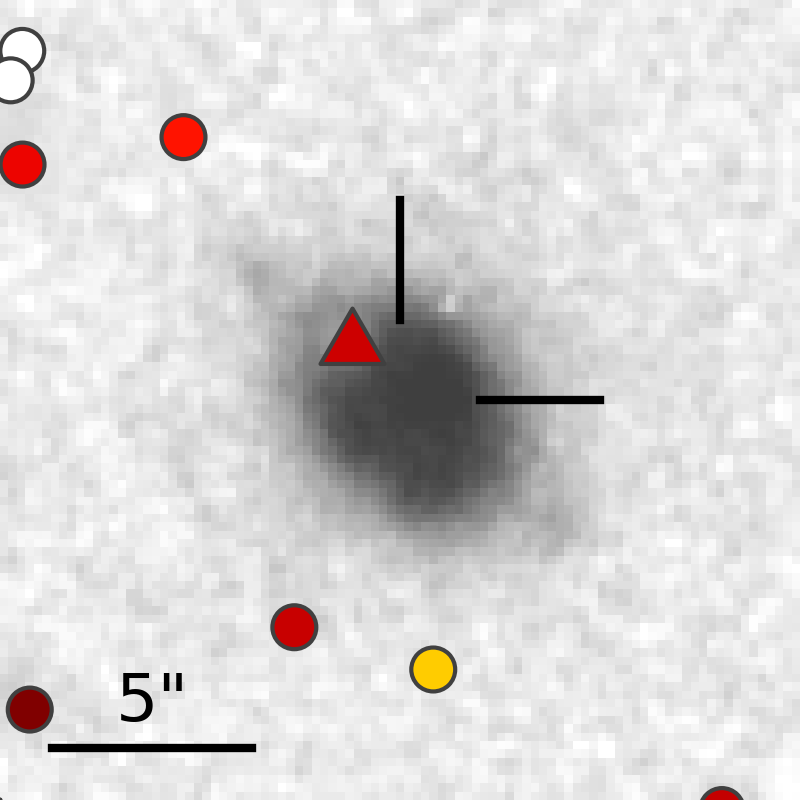}
        \enskip
        \includegraphics[width=0.45\hsize]{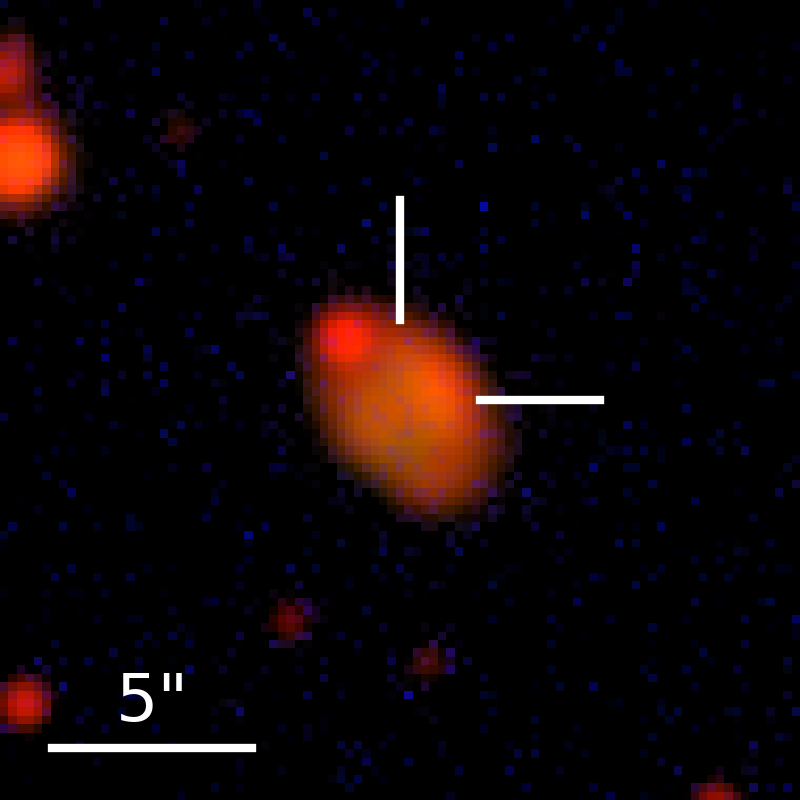}
        \caption{\label{fig:kb18comparison:k313}K 3-13 (PN G034.0+02.2)}
    \end{subfigure}
    \begin{subfigure}[t]{0.33\hsize}
        \centering
        \includegraphics[width=0.45\hsize]{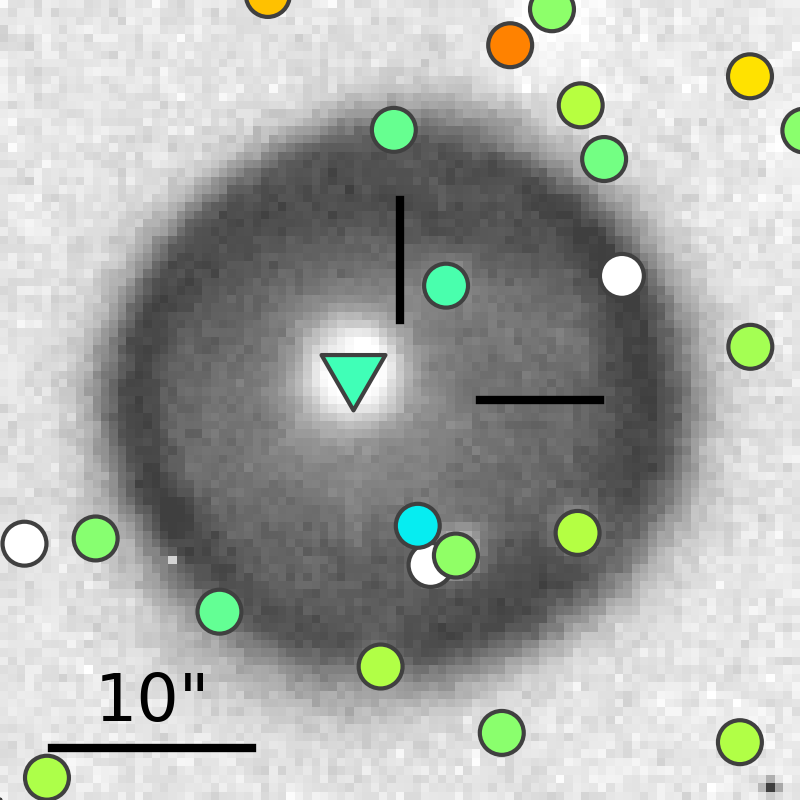}
        \enskip
        \includegraphics[width=0.45\hsize]{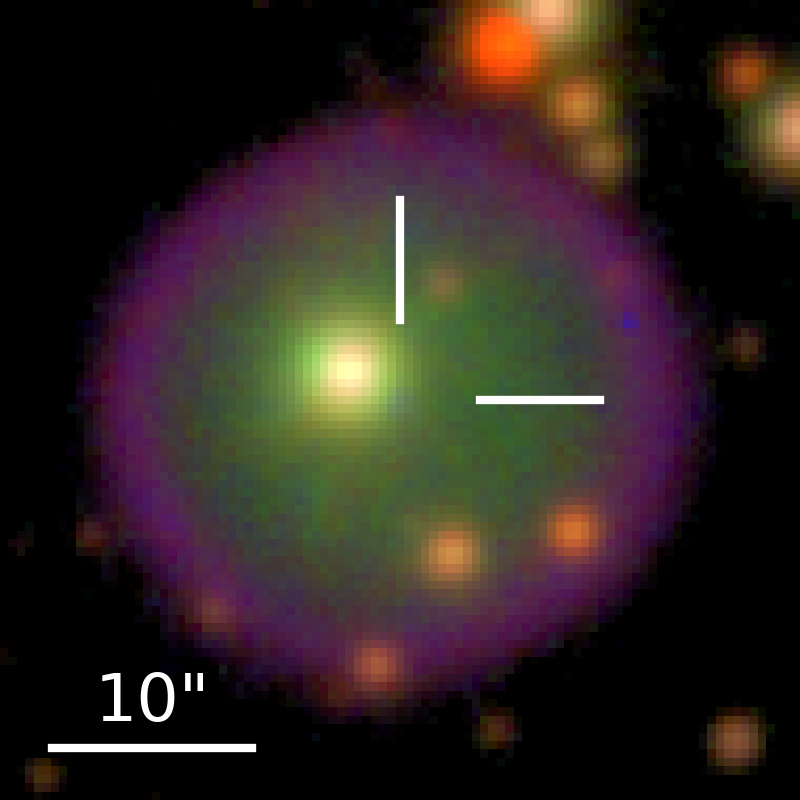}
        \caption{\label{fig:kb18comparison:lotr3}LoTr 3 (PN G265.1-04.2)}
    \end{subfigure}
    \par\medskip
    \begin{subfigure}[t]{0.33\hsize}
        \centering
        \includegraphics[width=0.45\hsize]{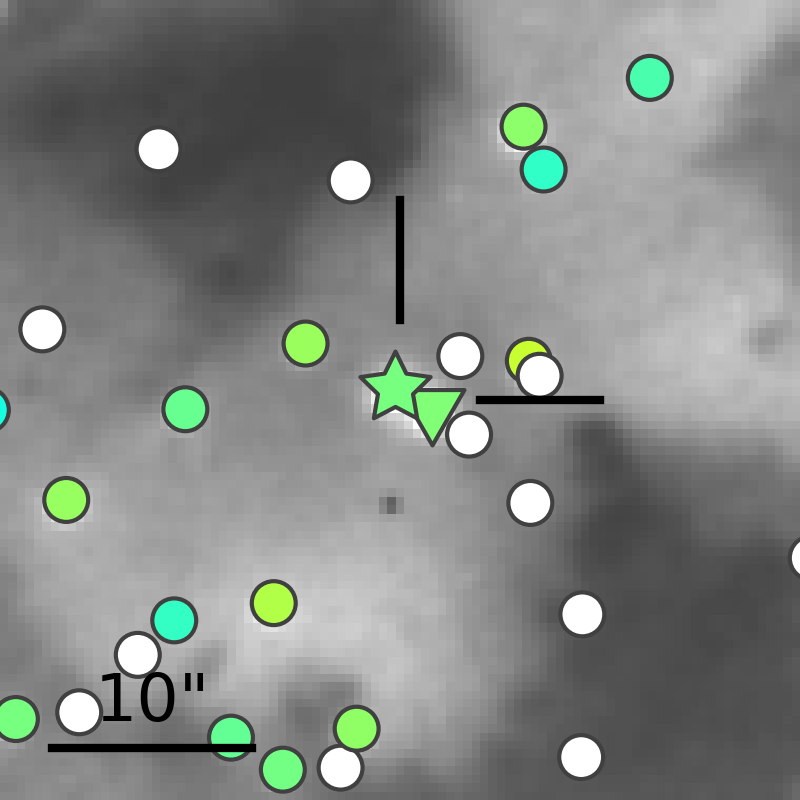}
        \enskip
        \includegraphics[width=0.45\hsize]{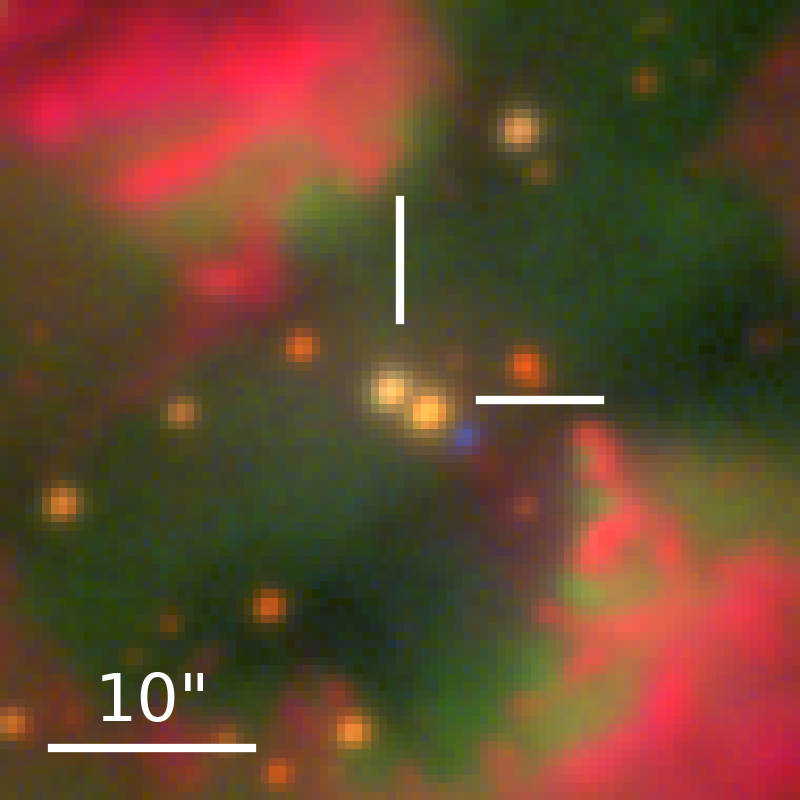}
        \caption{\label{fig:kb18comparison:ngc2899}NGC 2899 (PN G277.1-03.8)}
    \end{subfigure}
    \begin{subfigure}[t]{0.33\hsize}
        \centering
        \includegraphics[width=0.45\hsize]{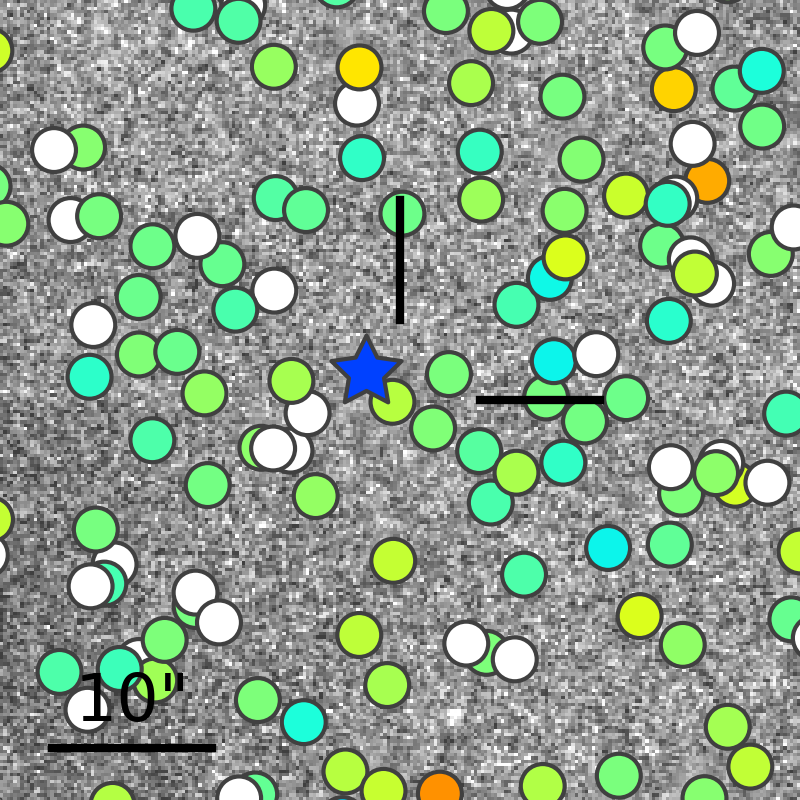}
        \enskip
        \includegraphics[width=0.45\hsize]{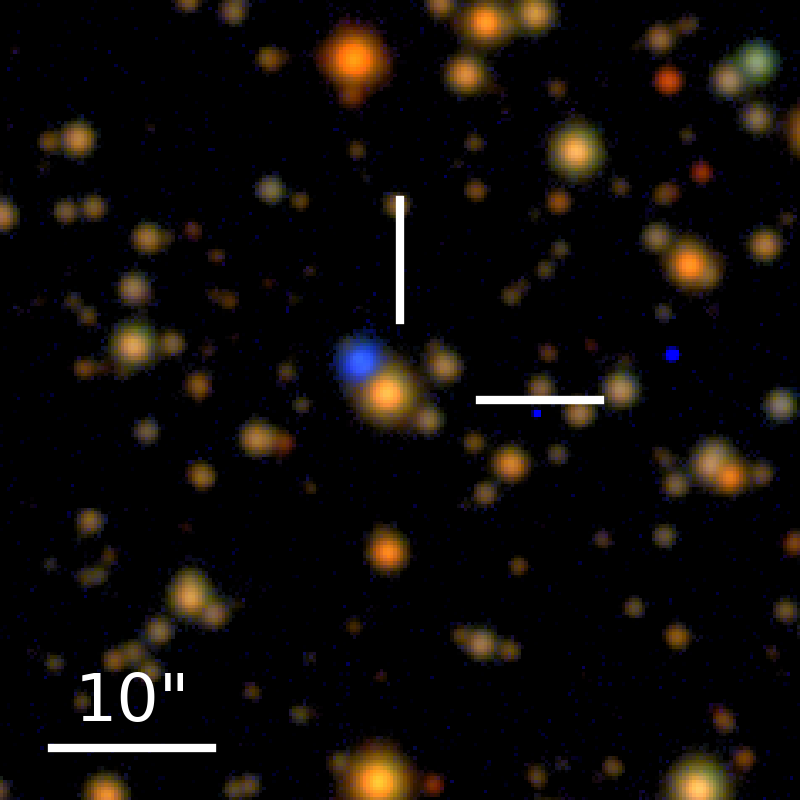}
        \caption{\label{fig:kb18comparison:sb39}SB 39 (PN G353.3-08.3)}
    \end{subfigure}
    \begin{subfigure}[t]{0.33\hsize}
        \centering
        \includegraphics[width=0.9\hsize]{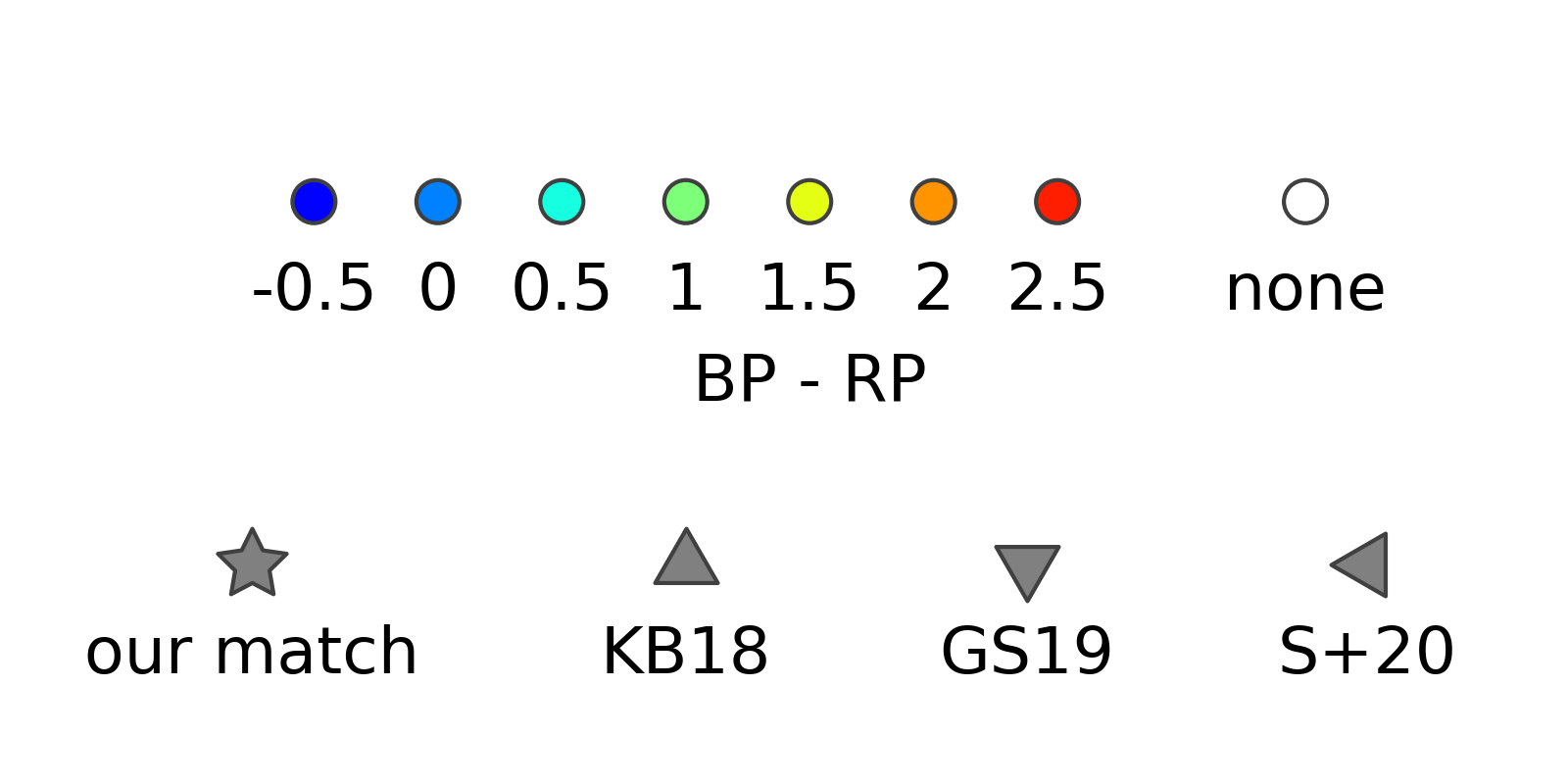}
    \end{subfigure}
    \caption{Paired quotient ($r^\prime$ -- H$\alpha$) and colour ($u^\prime$, $g^\prime$, $r^\prime$) images from VPHAS+ of selected PNe centred on their coordinates from HASH. North is up and east is to the left. The coloured markers overlayed on the quotient images show \emph{Gaia} detections with colour corresponding to BP -- RP as shown in the legend and shapes indicating matches from this and previous works. 
    The broadband colour images (with colours derived following \citet{rgbimageslupton}) are useful for comparison as they better capture the range of stellar colours and highlight blue central stars.
    }
    \label{fig:kb18comparison}
\end{figure*}

\subsubsection{Results}

The results of our comparisons are shown in Fig. \ref{fig:kb18reliability} and Table \ref{countstable}, with individual examples in Fig. \ref{fig:kb18comparison}. There are two considerations: whether the best match found by our method is the same as that found by a previous work, and the reliability assigned to that match by our method (and to the previous work's match, if different).

Following the thresholds defined earlier, we consider our method to agree with a previous work if it assigned reliability > 0.8 to that work's match, and to have rejected the previous work's match if it assigns reliability < 0.2. Rejections in which our method identified the same best match but assigned it low reliability indicate that our method considered there to be no good candidates (objects on the lower left diagonals of the scatter plots in Fig. \ref{fig:kb18reliability}). These low reliability candidates are excluded from our analysis and our published best matches catalogue. Rejections in which our method found a different best match appear as off-diagonal symbols in the scatter plots in Fig. \ref{fig:kb18reliability}; those cases in which the different best match is plausible (reliability > 0.2) count towards disagreements in the histograms in that figure as well as in the counts in Table \ref{countstable}.

We found good agreement overall with KB18, though there is a non-negligible fraction of disagreements (e.g. Fig. \ref{fig:kb18comparison}\subref{fig:kb18comparison:pb1}) and rejections (e.g. Fig. \ref{fig:kb18comparison}\subref{fig:kb18comparison:k313}). Many appear to be due to differences in input catalogue positions; manual spot-checks suggest that the positions in HASH are generally better than those in SIMBAD. 

We do believe that any potentially incorrect associations in KB18 are unlikely to substantially change the results of their distance comparisons, because of restrictions on colour and the outlier cuts that they used in their regressions. Indeed, our method matched all of their colour-selected sample (-0.65 < BP--RP < -0.25), and rejected only three of their sources with relative parallax errors better than 15\% (though these do not correspond to the outliers noted in their regression). Those rejected sources were those for K 3-55 (PN G069.7+00.0), for which we found no good match, and Hen 2-114 (PN G318.3-02.0) and Sh 2-71 (PN G035.9-01.1), for which we found different matches. In the latter two cases, the parallaxes of our matches had larger uncertainties and would not have been included in the KB18 regression had they been matched. However, while there was tension between the parallaxes of the sources in KB18 and the statistical distances to those PNe, with our matches that is no longer the case.

We also found many matches that were missed by KB18, matching 38\% of the PNe in SH10 and not already in KB18. This is slightly lower than our overall rate of 44\%. These new matches from SH10 that are not in KB18 tend to lack secure colour information, either missing BP--RP colours altogether or having a high BP/RP excess factor due to nebular contamination or crowded fields.\footnote{Examples are M 1-18 (PN G231.4+04.3) and Mz 1 (PN G322.4-02.6) with high excess factors, and Hen 1-6 (PN G065.2-05.6) and K 1-4 (PN G001.0+01.9) with no colour.} Some are also for compact PNe where the central star is likely not visible through the nebula.\footnote{An example is KFL 19 (PN G003.3-07.5).}

The catalogues of GS19 and S+20 have lower coincidence with ours than KB18, perhaps because they relied on purely positional cross-matching. As with KB18 there is a mix of rejected sources and disagreements as to which source is the best candidate.

Many of the sources in S+20 that disagree with ours have coordinates matching those of K03. Some of the central star positions in K03 appear to be those of field stars, sufficiently central to find the PN but otherwise unrelated.\footnote{An example is Abell 67 (PN G043.5-13.4). Our method found a faint blue source 1.7\arcsec away from the HASH position, while the position from K03 used by S+20 corresponds to a redder star 2 magnitudes brighter and offset from the HASH position by 7.1\arcsec.} The single band imagery used would have made it difficult to distinguish true central stars based on colour. Likewise, without narrowband imagery, stellar-like PNe are prone to confusion with stellar sources. The H$\alpha$ imagery in more recent surveys that the HASH positions are based on results in fewer misidentifications.

Even where input catalogue positions agree, our positional matching is more conservative than previous works (e.g. Fig. \ref{fig:kb18comparison}\subref{fig:kb18comparison:lotr3}). Most sources further than 1\arcsec are rejected by our method unless they are particularly blue (Fig. \ref{fig:results}, upper right). This contrasts with the 2\arcsec and 5\arcsec search radii used by GS19.

Spurious matches in both of these works tend to have lower distances than the PNe to which they are incorrectly associated, and to have relatively good parallax uncertainties that increase their chances of passing parallax quality cuts and making their way into scientific analysis. In the course of their H$\beta$ surface brightness distance scale calibration, S+20 found several objects in their sample with small physical radii (derived from the \emph{Gaia} parallaxes of their matches) that were noticeably poor fits to the overall trend. They termed these "low ionised mass" PNe, and excluded them from much of their analysis. We believe that most of this population is in fact explained by these PNe incorrectly being matched to nearby field stars, causing their distances to be underestimated. This in turn led to the physical radii being underestimated, as well as their ionised masses (ionised mass being proportional to flux divided by distance). Indeed, among the 18 objects marked as "low ionised mass" in S+20's Figure 3, 12 were rejected in our matching.

In general, even mismatches that do not result in obvious outliers can affect results, for example skewing distance scale calibrations or causing the local population of PNe to be overestimated.

\subsubsection{Individual objects}

Illustrative examples of individual objects are shown in Fig. \ref{fig:kb18comparison}. They have been selected to show different match scenarios and disagreements with previous works. For simplicity we have limited these examples to ones with imagery from VPHAS+, which restricts the examples to PNe visible from the South and located near the galactic plane. The VPHAS+ survey is particularly useful in that it includes the $u^\prime$ band, which can more clearly separate hot CSPNe from main sequence stars. It is important to note that these images are for illustrative purposes only, and that our method relied solely on the photometry and positions from \emph{Gaia}.

Some central stars have blue \emph{Gaia} colours, though many are missed by previous works relying purely on positional cross-matching:
\begin{itemize}
    \item \emph{M 1-18}. This is a straightforward case with a blue centrally located star. The PN is in the SH10 catalogue but its central star was missed by KB18 and other previous works, so it does not appear in the scatter plots of Fig. \ref{fig:kb18reliability}.
    \item \emph{PB 1}. We believe the redder, less central star selected by KB18 to be a misidentification.
    \item \emph{IC 1295}. The blue central star is 2.1\arcsec from the HASH position and is correctly selected by our method despite its separation. S+20 has selected a source with no \emph{Gaia} colour that is 3.2\arcsec away from the HASH position.
    \item \emph{SB 38}. The blue central star is 0.5\arcsec from the HASH position. GS19 has selected a source 2\arcsec away that is nearly 3 magnitudes brighter and appears to correspond to the PN's position in K03.
\end{itemize}
\noindent Other stars do not appear blue in \emph{Gaia}, for various reasons:
\begin{itemize}
    \item \emph{Hen 2-39}. The central star of this PN is dominated by a binary companion \citep{hen239binary} and appears red in \emph{Gaia}. Our method matched it based on its very central position, 0.2\arcsec from that in HASH.
    \item \emph{H 2-41}. The central star of this PN has no \emph{Gaia} colour and was matched by our method, albeit with low certainty, purely based on its position 0.6\arcsec from that in HASH. It is only through comparison with the VPHAS+ imagery that it can be seen to be the correct match, rather than the redder off-centre star selected by S+20.
\end{itemize}
\noindent Colour is less indicative for compact or stellar-like PNe, so accurate catalogue positions are particularly important:
\begin{itemize}
    \item \emph{M 1-38}. The central star of this PN appears to be obscured by the nebula itself, but there is a \emph{Gaia} detection.
    \item \emph{K 3-13}. No central star or bright spot is visible, and KB18's match is rejected by our method as too far away.
\end{itemize}
\noindent Colour is also important for disambiguation when separations are large or there are multiple possible options:
\begin{itemize}
    \item \emph{LoTr 3}. There appears to be no central star detection (nor is one visible in the imagery), and the purely positional cross-matches of GS19 and S+20 have selected a nearby source whose colour and absolute magnitude are consistent with a main sequence star. 
    \item \emph{NGC 2899}. The stars selected by GS19 and by our method both appear to be incorrect. The true central star is a fainter blue star visible towards the southwest, but it lacks a \emph{Gaia} colour and is therefore challenging to identify from the \emph{Gaia} data alone.
    \item \emph{SB 39}. As with NGC 2899, the true central star of this PN is not one of the sources nearest to its centre. However its blue colour in \emph{Gaia} has allowed our method to correctly identify it. It is not included in previous works, so it does not appear in the scatter plots of Fig. \ref{fig:kb18reliability}.
\end{itemize}

While this section has focused on individual objects in the context of matching, a small number of additional individual objects are discussed in Sect. \ref{tracks_discussion} and Sect. \ref{statistical_results} in the context of applications.

\subsection{Catalogue}
Our full best matches catalogue (Table B.1) is available online at the CDS, containing the highest reliability matches from \emph{Gaia} DR2 for all true, likely, and possible PNe in HASH, for which the reliability is at least 0.2 (our threshold for a possible match). The catalogue contains, for each PN, the PN G identifier and name, the \emph{Gaia} DR2 source ID for the single best match for that PN, and the reliability of that match determined by our algorithm. The reliability should be used as a filter to limit analysis to high confidence matches (e.g. reliability > 0.8), with the particular threshold being dependent on the application. In addition the table contains selected columns from the \emph{Gaia} catalogue and from HASH are that are particularly relevant to the matching and to the science results presented in this work. From \emph{Gaia} we include the position, colour, magnitude, and parallax of the best matching sources, as well as a derived quantity indicating the astrometric goodness of fit. From HASH we include PN positions and radii, and their confirmation status (confirmed, likely, or possible PN). We also include the separation between the PN and \emph{Gaia} source positions. Finally, for convenience, we include cross-match flags with the previous works mentioned in this section and the list of PN binaries in Fig. \ref{fig:reddening}.

It is important to treat the matches that we provide probabilistically and appropriately in context. There is a lot of additional information not used in our matching, such as source magnitude and parallax, that can help to disambiguate uncertain cases. For example, if a candidate source has a precise parallax that strongly disagrees with other reliable distance measures or leads to an implausible physical nebula size, that adds weight to the source in fact being coincidental, especially in the absence of strong evidence from other non-positional features such as colour. These caveats are, of course, not unique to our work, though ours is the first to attempt to quantify the uncertainties involved.

\section{Applications} \label{applications}

The subset of matched \emph{Gaia} sources with parallaxes offers a significant increase in the number of primary galactic PN distance measurements, even with additional restrictions on parallax uncertainties or other quality indicators (e.g. Fig. \ref{fig:parallaxerrors}). We present some indicative results using these parallaxes to characterise PN physical properties and revisit the statistical distance scale of FPB16.

Throughout this section we use $\omega$ to indicate parallax, consistent with the notation in previous works based on \emph{Gaia} data (e.g. \citet{gaiadr1astrometry}).

\subsection{Physical parameters}

Accurate distance measurements enable us to transform angular sizes of PNe to physical radii. Combined with kinematical assumptions physical radii can determine the age of the nebula. Moreover we can also determine the luminosity of the central star, which is also related to its age and thereby its position on the evolutionary track between AGB and White Dwarf stages.

\begin{figure}
    \centering
    \includegraphics[width=\hsize]{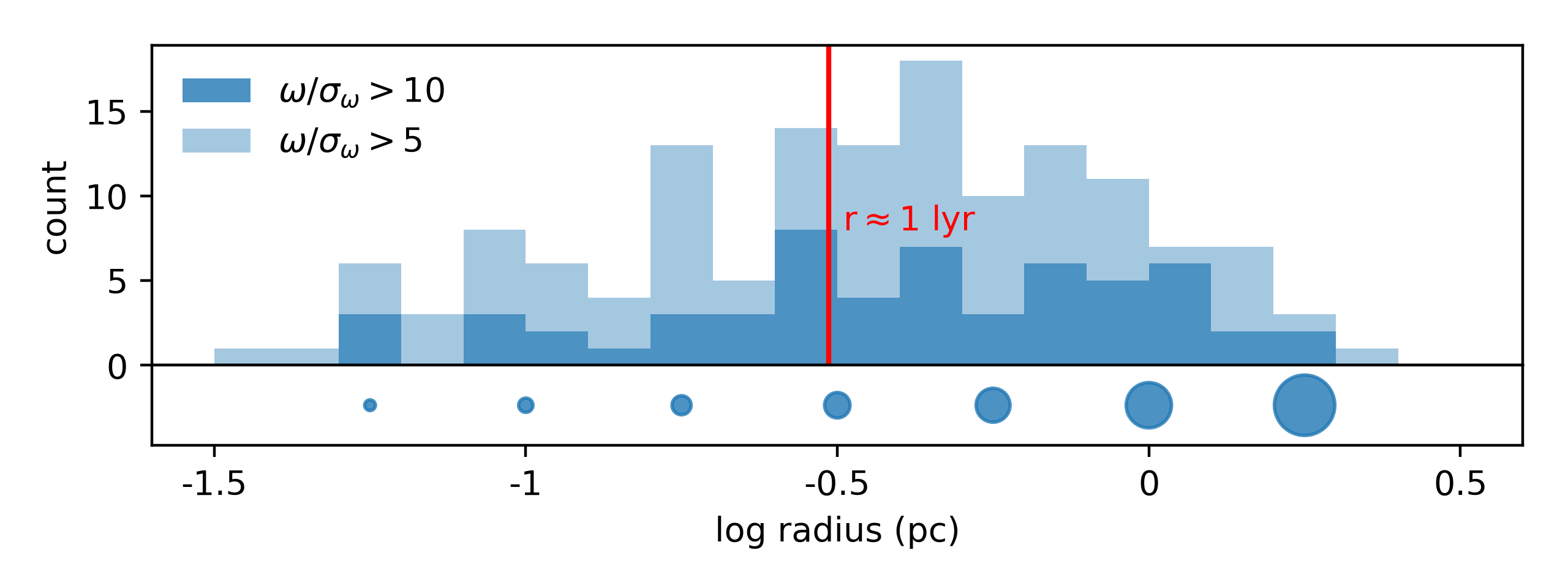}
    \caption{Histogram of PN physical radii derived from \emph{Gaia} parallaxes of matched CSPNe with various relative parallax error cutoffs. For comparison with Fig. \ref{fig:hrdiagram}, circle sizes used to denote physical radii are shown in the lower panel.}
    \label{fig:radii}
\end{figure}

The distribution of physical radii is shown in Fig. \ref{fig:radii} for PNe whose matched central stars have a relative parallax error better than 20\%. With these errors parallax inversion produces relatively well-behaved distance estimations \citep{bailorjonesparallax}, which we deem acceptable for the indicative results that we present, particularly since we are not making overall population characterisations that would be biased by this sort of selection.

\begin{figure}
    \centering
    \includegraphics[width=\hsize]{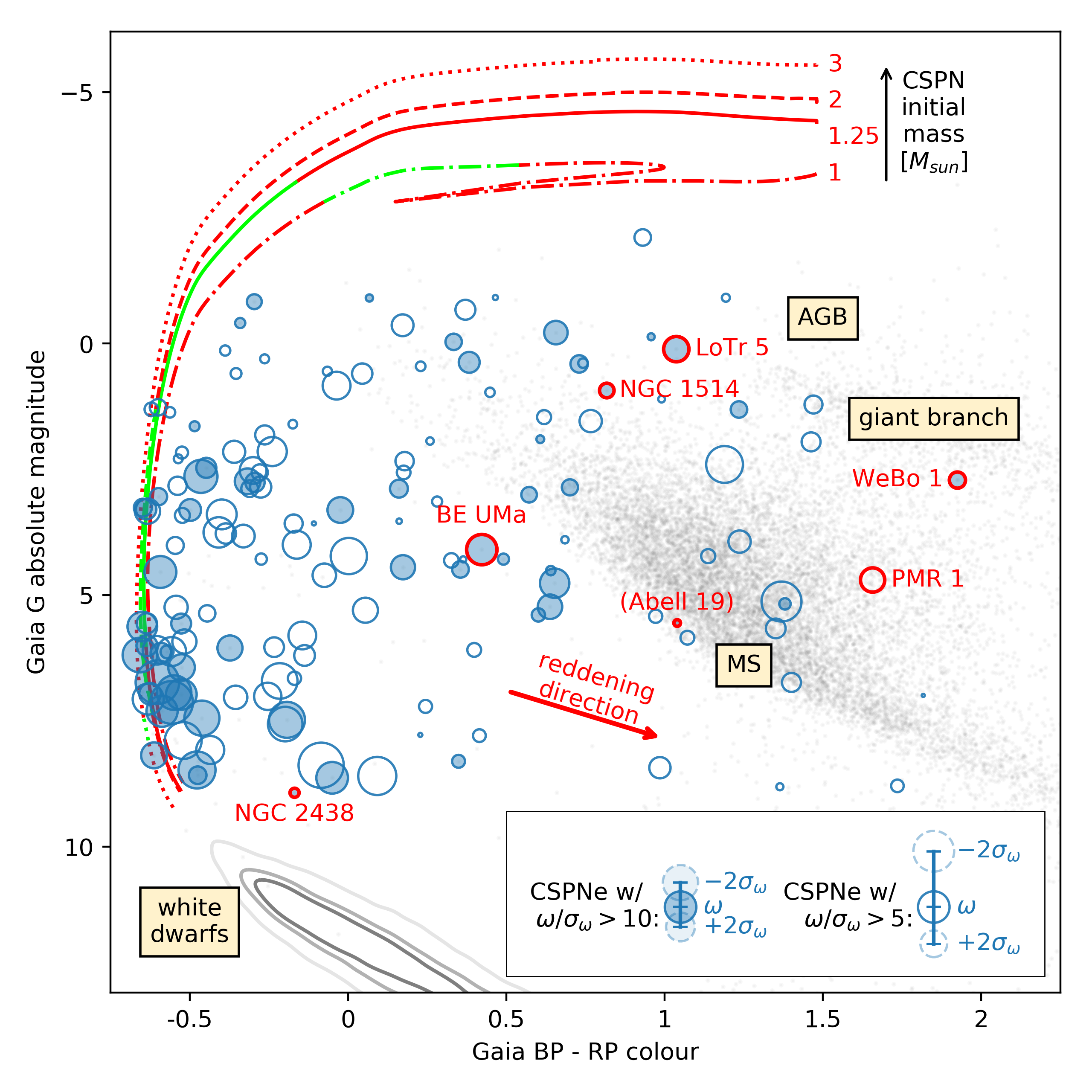}
    \caption{PN central stars plotted on an observational HR diagram, with the circular markers scaled according to the physical radii of the PNe as in Fig. \ref{fig:radii}. Filled circles indicate objects with the lowest uncertainties. Individual PNe referenced in the text are coloured red rather than blue and accompanied by the PN name. Red lines represent CSPN tracks from \citet{bertalomicspntracks} for solar metallicity and various initial masses, with the green portions of the line denoting time since leaving the AGB of between 1000 and 20000 years, indicative of the sorts of timescales during which a PN could be visible. The peak temperatures of these tracks, through which the stars evolve relatively quickly, are located at an absolute \emph{Gaia} magnitude around 5 (see text for details). In the background, the grey points are the other sources that were loaded in the 60\arcsec search windows, with $\sigma_\omega/\omega < 10\%$. They trace out the main sequence (MS) and giant branch. The beginning of the AGB is also labeled, with its position taken from \citet{gaiahrdiagram}. White dwarfs are shown separately, as they are too rare to appear otherwise, with the grey contours in the lower left representing the 10, 30, and 50\% density contours of the observed high confidence white dwarf candidates from \citet{gaiawhitedwarfcandidates}, where the same quality cuts have been made as for the background points.}
    \label{fig:hrdiagram}
\end{figure}

For central stars in \emph{Gaia} with both full astrometric solutions and full photometry, we can combine the \emph{Gaia} G band magnitude and the distance estimate from the parallax to estimate the absolute magnitude, and plot this against BP -- RP colour in an observational Hertzsprung–Russell diagram (HRD), following \citet{gaiahrdiagram} (Fig. \ref{fig:hrdiagram}). Even without correction for reddening and extinction, most of our matches occupy an otherwise sparsely populated region of the HRD, bluer than the main sequence and giant branch but also brighter than white dwarfs.

\subsubsection{Theoretical tracks}

Comparison to theoretical tracks requires mapping between the physical stellar parameters of effective temperature $T_\textup{eff}$ and bolometric luminosity $L$ and the observed \emph{Gaia} BP -- RP colours and G magnitude.

The goal of the \emph{Gaia} astrophysical parameters inference system (Apsis) is to perform the mapping starting from \emph{Gaia} observations, deriving the mapping based on machine learning techniques \citep{gaiaapsisprelaunch}. It will ultimately use the \emph{Gaia} spectrophotometry and account for reddening and extinction as well. In \emph{Gaia} DR2, temperatures are available for less than 10\% of sources, and luminosities for less than half of those. Moreover, because of the limited temperature range of the training data, the $T_\textup{eff}$ values that the model does produce do not go above 10000 K \citep{gaiadr2apsis}, making them unhelpful for the much higher temperature range expected for CSPNe.

Instead we perform the mapping in the opposite direction, transforming physical parameters into observables. Such transformations were provided pre-launch for main sequence and giant stars by \citet{jordigaiaphotometry}, and for white dwarfs in a followup paper by \citet{gaiawdphotometry}. Though the latter transformations cover higher surface gravities and extend the range of effective temperatures, the two together still miss most of the CSPN evolutionary tracks that we wish to cover. Fortunately, the transformation for higher temperature objects such as CSPNe is largely independent of metallicity and surface gravity. We used the revised BP and RP passbands from \citet{gaiaphotometry} and assumed blackbody spectral energy distributions (SEDs) to generate expected BP -- RP colours. We found that these were bluer than expected from the pre-launch papers and speculate that this is due to higher sensitivity than expected at the shortest wavelengths. The bluer tracks are also a better fit for the observed data. For transforming luminosity into absolute \emph{Gaia} G magnitude we adopt the bolometric corrections from \citet{gaiawdphotometry} for a surface gravity (log \emph{g}) of 7, fitting a cubic spline to extrapolate to the highest temperature regime. The G passband more closely matches the nominal, pre-launch passband, so we do not expect these calculations to change significantly.

Using these transformations, we plot a selection of tracks from \citet{bertalomicspntracks} for solar metallicity (Z$_0$ = 0.01, versus 0.0134 for the sun) and a range of initial masses. Bolometric corrections change the shape of the tracks from those in the temperature versus luminosity space, with higher temperature objects at the same luminosity having more of their flux at ultraviolet wavelengths outside the \emph{Gaia} G band. Thus peak temperature occurs at a G absolute magnitude of around 5, with higher temperatures appearing fainter.

The theoretical tracks are relatively close to each other in the BP -- RP colour space, as the \emph{Gaia} BP -- RP colours are not highly sensitive in the high temperature regime occupied by CSPNe. Between this and the degeneracy between temperature and reddening, we are not able to constrain initial masses and ages from the \emph{Gaia} DR2 photometry alone (this degeneracy and the insufficiency of \emph{Gaia} photometry alone was noted in for white dwarfs in \citet{gaiawhitedwarfcandidates}). Such determinations require additional photometry or spectroscopy to better constrain and disentangle reddening and temperature. The \emph{Gaia} estimated distances combined with dust maps may prove useful in this regard.

\subsubsection{Discussion} \label{tracks_discussion}

\begin{figure}
    \centering
    \includegraphics[width=\hsize]{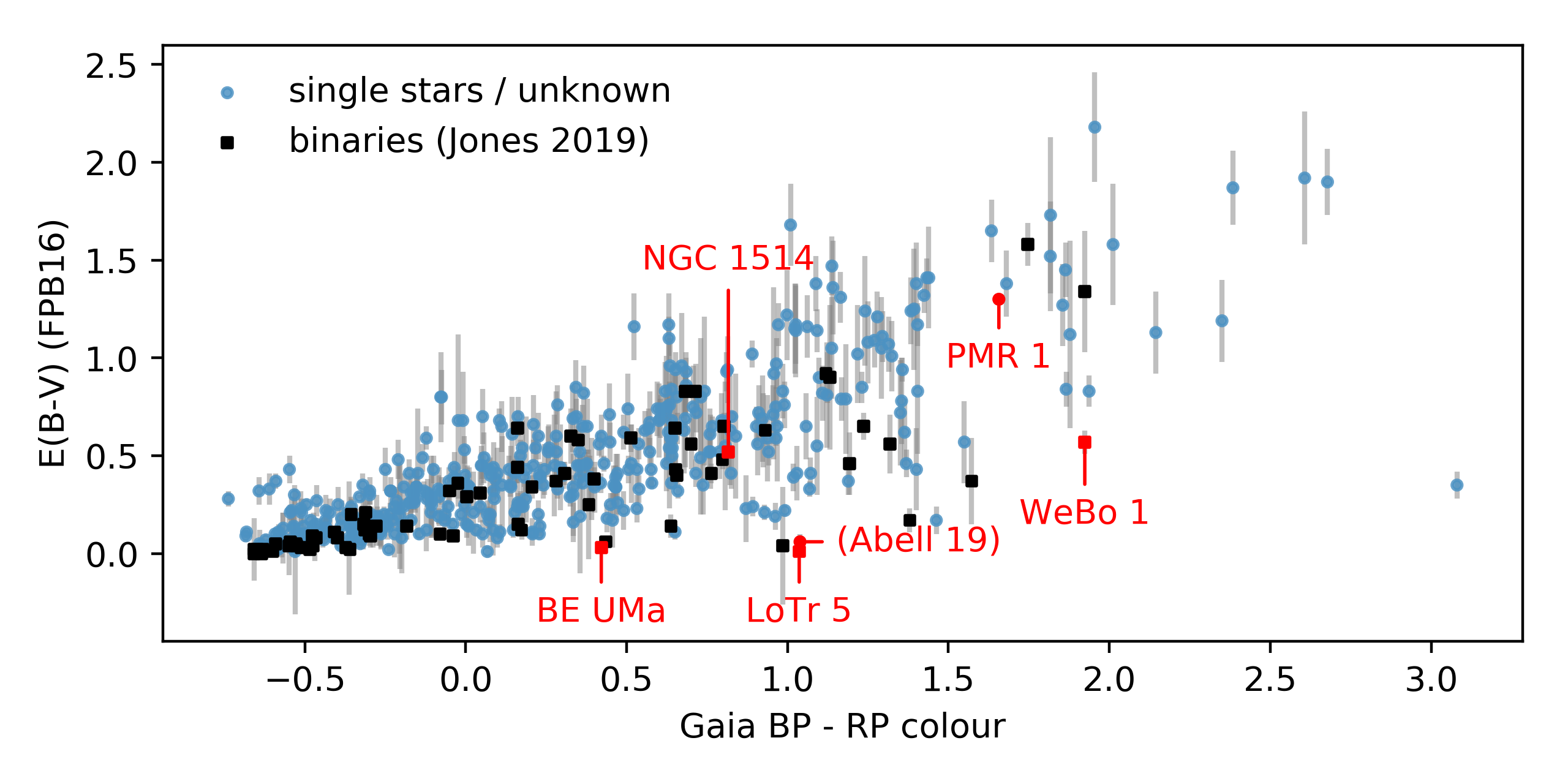}
    \caption[]{Reddening values E(B-V) and their given uncertainties taken from FPB16's statistical distance compilation\footnotemark{} plotted against \emph{Gaia} BP -- RP colours for all matches with reliability > 0.8 (not limited by parallax uncertainties). Known and suspected binary systems taken from the compilation of David Jones\footnotemark{} are highlighted as black squares.\footnotemark{} Objects lying below the trend (objects appearing red in \emph{Gaia} with low reddening) could be binary systems or have significant reddening internal to the nebula, or could have dubious identification. Relevant individual objects mentioned in the text are shown in red.}
    \label{fig:reddening}
\end{figure}
\addtocounter{footnote}{-2}
\footnotetext{Reddening for PMR 1 is taken from \citep{pmr1wolfrayet} as it is missing from FPB16. It is missing error bars because there was no uncertainty published for the estimate.}
\stepcounter{footnote}
\footnotetext{\url{http://www.drdjones.net/bcspn/}, as of March 11, 2019.}
\stepcounter{footnote}
\footnotetext{PHR J1510-6754 (PN G315.4-08.4) is also included as a binary as noted by \citet{hillwigbinary}; it is not included in Jones' list but the reference was found incidentally in a cursory literature search for objects lying below the trend.}

We find that most of our matched central star colours and absolute magnitudes are well explained by the theoretical tracks plus reddening effects, and that the physical sizes of the nebulae are consistent with the evolutionary direction of their central stars (in that younger and therefore brighter central stars have less evolved nebulae).

Several of the CSPNe are inconsistent with the evolutionary tracks; that is those with relatively red BP -- RP colours whose de-reddened projection onto the theoretical tracks is a poor fit. We focus on those with that are for resolved nebulae (so that the \emph{Gaia} detection is of the central star rather than the nebula itself) and have low BP/RP excess factors (indicating well-behaved photometry; colour uncertainty from high excess factors likely dominates any flux uncertainties).

One explanation is that these are binary systems where the light from actual progenitor of the PN is dominated by a main sequence companion. A few examples that we checked are LoTr 5 (PN G339.9+88.4) and BE UMa / LTNF 1 (PN G144.8+65.8), which have the largest absolute latitudes of the red sample, meaning that their colours are less likely to be reddened. Both of these are in fact known binary systems (\citet{joneslotr5ngc1514binary} and \citet{fergusonbeumbinary} respectively), as is NGC 1514 (PN G165.5-15.2) in the first reference. WeBo 1 (PN G135.6+01.0), the reddest star in the sample with a large BP--RP value of 1.9, is also a binary \citep{webo1binary}, while the second reddest star, the central star of PMR 1 (PN G272.8+01.0) with BP--RP equal to 1.7, is noted in the literature to simply be heavily reddened \citep{pmr1wolfrayet}. These are highlighted in Fig. \ref{fig:hrdiagram} and Fig. \ref{fig:reddening} along with the other individual PNe mentioned in this section.

One star that does appear inconsistent with the nebula size versus absolute magnitude trend is NGC 2438 (PN G231.8+04.1). The veracity of its identification as the central star is confirmed by its colour, but its parallax measurement places it around 422pc away, with less than 10\% error, which is inconsistent with other distance determinations for the PN. The statistical scale of FPB16 estimates the PN's distance to be 1.54kpc $\pm$ 0.44kpc, consistent with an even further away estimate from central star modelling that was used as part of that paper's calibration. We believe that the parallax errors in this case may not be well behaved, supported by the source having high astrometric excess noise (and renormalised unit weight error (RUWE) = 2.39). Assuming NGC 2438 to be further away removes tensions, leading to a larger physical nebula size and a brighter central star.

The HRD suggests the possibility of further refinement of the matching itself based on parallaxes, with the derived absolute magnitudes disambiguating between possibly reddened CSPNe and main sequence stars, and the derived distances allowing for the calculation of a physical radius which can then be checked for compatibility with other knowledge about the PN. Stars that do lie on the main sequence and have plausible parallaxes merit further investigation as possible mismatches, binaries, or reddened single central stars.

An example of an implausible match is that of the "likely" PN Abell 19 (PN G200.7+08.4). The centrally located star is nearby, at around 250 pc away, with small parallax uncertainties and a colour and magnitude that place it neatly on the main sequence (Fig.  \ref{fig:hrdiagram}). While the colour could be explained by reddening, significant reddening is unlikely given the star's relatively close proximity; this is confirmed by its low estimated reddening in FPB16 (Fig. \ref{fig:reddening}). It is also unlikely to be a binary companion, because the PN physical radius of 0.05 pc (log physical radius of -1.3) resulting from its parallax is smaller than almost all of the PNe in our sample (Fig. \ref{fig:radii}) and consistent with a very young PN, which would then have a much brighter central star and nebula. Thus we can conclude that this candidate is more likely to be a nearby field star.

\subsection{Statistical distance scales}

The parallaxes from \emph{Gaia} also offer an opportunity to evaluate and ultimately refine statistical distance scales. We focus on the H$\alpha$ surface brightness to physical radius relation from FPB16, which was calibrated based on distances to galactic and extragalactic PNe derived from a variety of primary techniques, including parallaxes from the HST and USNO but not from \emph{Gaia}.

\begin{figure}
    \centering
    \includegraphics[width=\hsize]{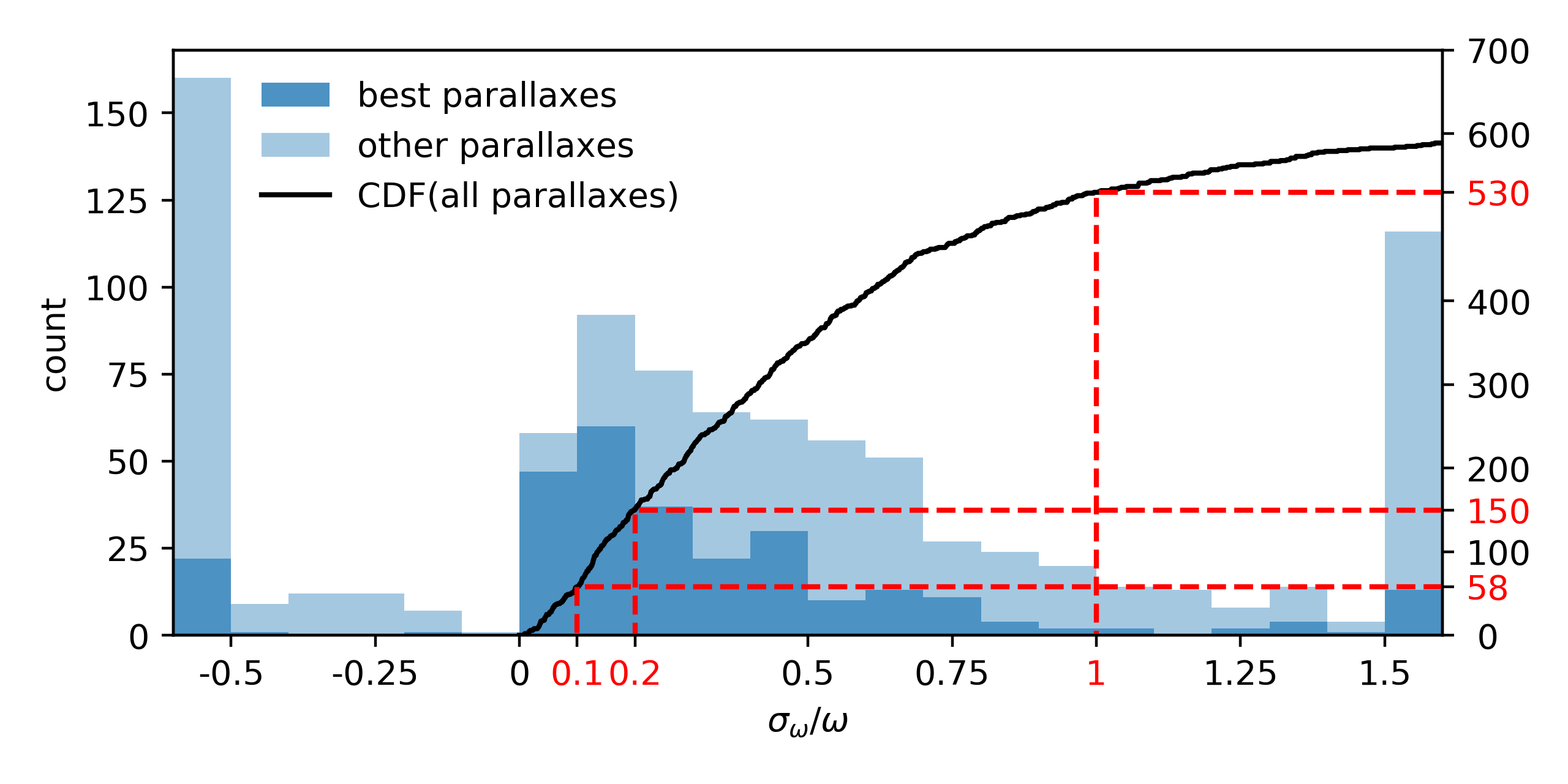}
    \caption{Relative parallax errors $\sigma_\omega/\omega$ for the best matches (reliability > 0.8) subsample of confirmed PNe, along with the cumulative counts below various reliability thresholds for positive parallaxes (in black). The bins at either end represent the counts or matches with $\sigma_\omega/\omega$ falling outside of the range (-0.5, 1.5). Within the sample, those parallaxes meeting more stringent criteria (reliability > 0.98, $\sigma_\omega$ < 0.2 mas, RUWE < 2, \texttt{visibility\_periods\_used} > 8) are indicated by the darker shaded area of the histogram. This subset is used for the \citet{frewsurfacebrightness2016} distance comparisons.
    }
    \label{fig:parallaxerrors}
\end{figure}

\subsubsection{Distance ratios}

The caveats present in using parallaxes to estimate distances are well known \citep{xlurigaiaparallaxes}; in particular naive parallax inversion does not produce a statistically sound distance estimate for any reasonable choice of prior, and any attempt to limit an analysis to parallaxes with relative error $\sigma_\omega/\omega$ below some threshold (as was done in the previous section), or even to positive parallaxes (as inverting negative parallaxes is unphysical) introduces biases. We can avoid these caveats by staying in the space of parallaxes, where the errors are well behaved.

The notion of distance ratios described in \citet{smith2015} avoids these caveats by sidestepping parallax inversion entirely and taking the product over objects of measured parallaxes $\omega'$ and estimated statistical distances $d'_s$ to form a distance ratio
\begin{equation} \label{eq:distance_ratio}
    R_S = \omega' d'_s
\end{equation}
with associated uncertainty
\begin{equation} \label{eq:distance_ratio_error}
    \sigma^2_R = d^2_S\sigma^2_\omega + \omega^2\sigma^2_S + \sigma^2_\omega\sigma^2_S
\end{equation}
where $d_S$ is the true distance $d$ multiplied by the distance ratio $R_S$ and $\sigma_S$ is the standard error on the statistical distance.

The distance ratio can be used to measure both errors in the intercept of a statistical relation (through deviations in the mean distance ratio away from unity) and in its slope (through correlations between the distance ratio and the estimated physical radius or statistical distance).

\subsubsection{Methods}

We consider the set of confirmed PNe for which FPB16 published statistical distances (1024 in total) and for which we have found reliable central star matches with high quality parallax measurements.

To limit the effect of poorly behaved parallaxes we apply quality cuts similar to those used in \citet{gaiahrdiagram}. We require a slightly higher number of observations than the threshold for inclusion into the Gaia data release, that is $\texttt{visibility\_periods\_used} > 8$. Additionally we set an upper limit on the renormalised unit weight error (RUWE) of 1.4 as recommended by \citet{gaiaruwe}. This is a goodness-of-fit statistic that indicates how well the \emph{Gaia} astrometric solution matches that expected for a single star \citep{gaiaastrometry}. Finally, we apply a cut on the absolute uncertainty of the parallax itself, requiring $\sigma_\omega < 0.2$ mas.

The aim of these cuts is to reduce the overall uncertainty in the average distance ratio without biasing its value. Thus we choose a cut based on absolute rather than relative parallax error, which avoids truncation biases. As parallax uncertainty is related to apparent magnitude, any cut based on parallax does bias the selection towards brighter and nearer objects, but that is unavoidable. It is worthwhile to note that there is also a risk that the cut based on RUWE could bias the sample by preferentially eliminating binary systems, though we do not see strong evidence that this is the case.

To avoid the effect of any incorrect matches we also apply a stricter reliability cut of reliability > 0.98, which keeps the vast majority of the matches.

The quality cuts leave us with 160 objects out of the 636 objects from FPB16 that we matched.

For many PNe, FPB16 provided multiple distance estimates - one based on a general trend, and one based on a subtrend for PNe that are classified as either optically thick or optically thin. The subtrend relationships have different slopes from each other and lower scatter. The calibrating set in FPB16 was chosen to represent a range of PN properties, and is balanced between optically thin and thick objects. If the subset that we compare is a different mixture, it will deviate from the mean trend even if the distances are correct. We consider this in the following section.

\subsubsection{Results} \label{statistical_results}

\begin{figure}
\centering
\includegraphics[width=\hsize]{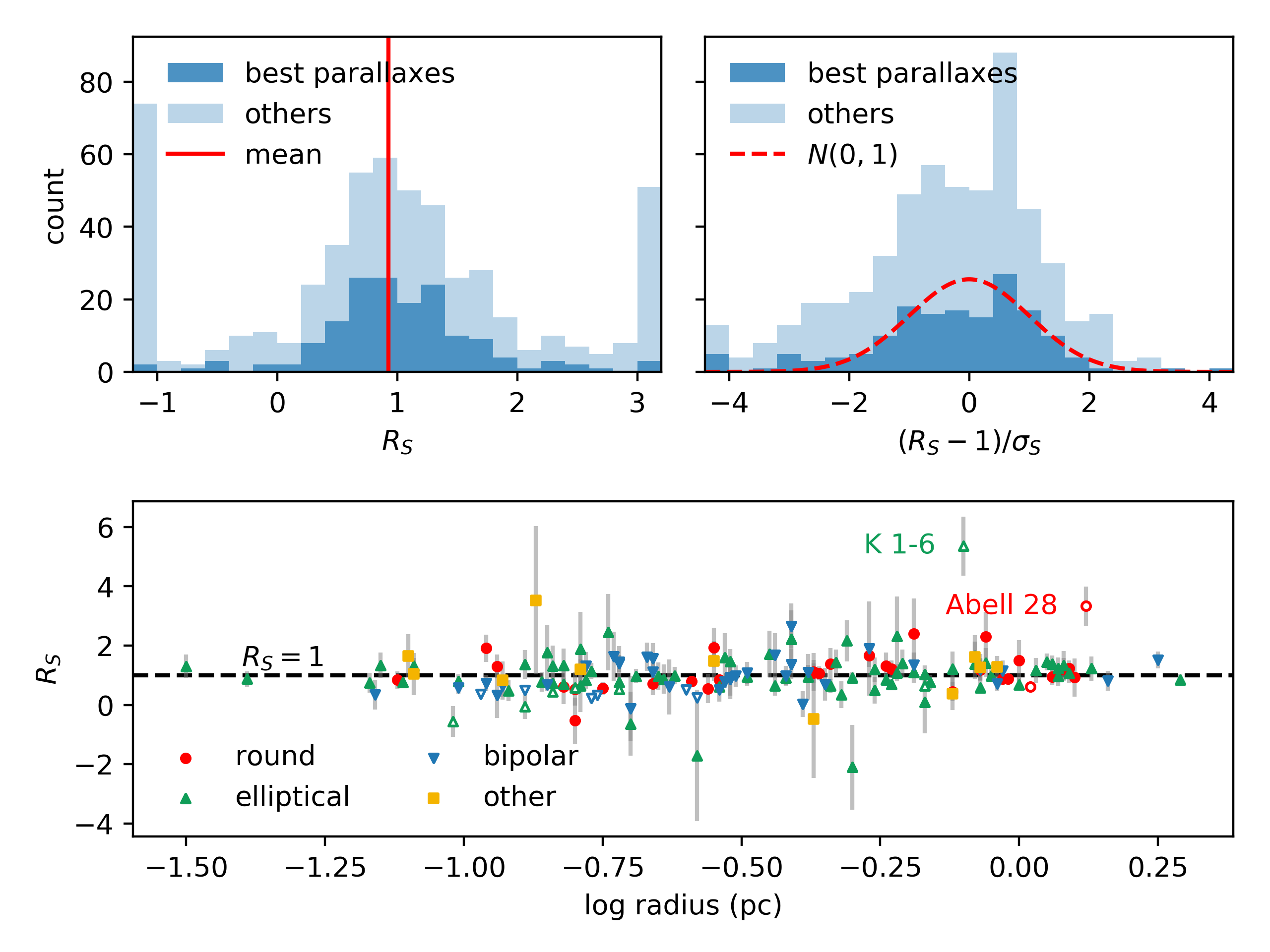}
  \caption{Histograms of distance ratios $R_S$ and normalised distance ratios $R_S / \sigma_S$derived from comparison between \emph{Gaia} parallaxes and statistical distances (using subtrends) from FPB16. Ratios are plotted for both the higher quality set of parallaxes (see text) and rejected parallaxes for comparison, in dark and light blue respectively. The plot on the left shows the raw distance ratios, with the mean value of 1.03 $\pm$ 0.06 for the best quality parallax set. On the right the distance ratios have been re-centred around $R_S=1$ and divided by their estimated uncertainties $\sigma_R$. Though the distribution of distance ratios is not expected to be Gaussian, a standard normal distribution is over-plotted for comparison. Below is a scatter plot depicting the distance ratios of the best parallax subset against the physical radius derived from the statistical distance. Marker colours and shapes show morphological classifications taken from HASH. Trends in this plot (that is, a correlation between distance ratio and radius) would be indicative of a slope differing from that derived in FPB16. Filled markers have $R_S$ within $2.5\sigma_S$ of 1. Outliers are empty markers, with the two outliers specifically mentioned in the text highlighted. The correlation coefficients are 0.18 and 0.08 with and without the outliers respectively. The former is very weakly significant, while the latter is not.
  }
   \label{fig:distanceratios}
\end{figure}

Using the mean trends gives a mean distance ratio of $1.15 \pm 0.07$ for the 160 objects that passed the quality cuts, while using the subtrends reduces this ratio to $1.03 \pm 0.06$ (the uncertainties are calculated via bootstrapping). We note that the matched PNe show a preference for optically thin PNe relative to the mixture of thick and thin PNe that formed the mean trend in FPB16, which could be due to optically thin PNe being more likely to have visible central stars. The subtrend in FPB16 for optically thin PNe has such PNe having lower surface brightnesses for the same physical radius, which translates to the mean trend overestimating the physical radii for these objects and thus overestimating their distances. This is consistent with the difference in mean distance ratios we see comparing the mean trend and subtrends.

The results using the subtrends are shown in Fig. \ref{fig:distanceratios}. On average the \emph{Gaia} parallaxes meeting our quality cuts are consistent with the associated FPB16 statistical distances. This is not surprising given the many extragalactic distances in the set of calibrating distances and the use of parallaxes in the calibration itself, which mean that the distance scale is unlikely to deviate from a true distances by the factors of two that older scales suffered from \citep{smith2015}. There is a slight suggestion of a dependency on physical radius but the uncertainties are too large to draw a meaningful conclusion. Grouping by morphology (lower half of Fig. \ref{fig:distanceratios}), we find no significant deviations from a mean distance ratio of unity, with round PNe having the largest deviation at $1.15 \pm 0.12$.

We see some notable outliers, objects for which $|(R_S - 1) / \sigma_S|$ is large. Only a couple have parallax-derive distances significantly smaller than their statistical ones:
\begin{itemize}
    \item \emph{K 1-6}. The statistical distances from both the mean trend (1.85 $\pm$ 0.53 kpc) and thin trend (1.45 $\pm$ 0.27 kpc) for K 1-6 (PNG 107.0+21.3) appear to be significant overestimates relative to its central star parallax, which places it within 500 pc. The distance from the thin trend is smaller and thus closer, but  This PN was studied in \citet{k1-6frew2011}, which noted tensions between different distance estimates for that nebula in terms of its surface brightness and the properties its binary central stars; they adopted a distance of 1kpc, halfway between FPB16's statistical distances and that suggested by the parallax from \emph{Gaia}. They also noted a range of possible distances based on the spectroscopic parallax of the binary central star companion, with the short end of those distances being consistent with the now observed trigonometric parallax.
    \item \emph{Abell 28}. Similarly to K 1-6, the \emph{Gaia} parallax for the blue central star of Abell 28 (PNG 158.8+37.1) places it within 500 pc, much closer than its statistical distances of 1.67 $\pm$ 0.48 kpc and 1.29 $\pm$ 0.25 kpc from the mean and thin trends respectively. The parallax-derived distance places it in the population of "subluminous" PNe noted in Sect. 4.3.4 of FPB16, with Abell 28 then occupying a place in the surface brightness versus physical radius plane near that of RWT 152 (PNG 219.2+07.5) (the parallax of the central star of RWT 152 itself is consistent with both the primary and statistical measurements).
\end{itemize}

On the other end of the scale there are several objects whose parallax-derived distances are significantly larger than their statistical ones (empty symbols below the dashed line in Fig. \ref{fig:distanceratios}). There is a suggestive excess of elliptical / bipolar objects in this set that would match the trend that FPB16 noted of bipolar objects having higher surface brightnesses, but even comparing the calibrating distances of those objects the \emph{Gaia} parallaxes shows significant disagreement by up to a factor of two, for example for Hen 2-11 (PN G259.1+00.9), whose parallax of 0.5 mas gives it a 2$\sigma$ distance range of 1.25 to 5 kpc from \emph{Gaia}, outside of the relatively confident 730 pc estimate derived from modelling of its binary star by \citet{hen211jonesboffin} that was also used in the calibrations by FPB16. One possibility is that the parallaxes are themselves skewed by binarity, as in \emph{Gaia} DR2 only single stars are modelled. Also, as the uncertainties in statistical distances are correlated with the statistical distances themselves, statistical distances that are underestimates also have underestimated uncertainties, which in turn means that the uncertainty in the distance ratio is underestimated. This effect was noted by \citet{smith2015}.

\subsubsection{Discussion}

The fraction of outliers would increase significantly if we lowered the reliability threshold of our method and accepted nearest neighbour \emph{Gaia} sources that we had not considered to be matches. The selection of such mismatches based on distance ratios is biased towards nearby objects, which tend to have lower parallax errors (on account of being brighter) and larger parallaxes that more tightly constrain their distances. Such mismatches will become more noticeable in future data releases as parallax uncertainties tighten, however, even mismatches that are individually consistent within errors will globally skew any calibration or evaluation, making it important to have a robust selection process to begin with. As with the HRD in the previous section, additional data, in this case distance priors based on statistical distances derived from nebula properties, can be used to further refine the matching by placing bounds on reasonable parallaxes.

Ultimately the \emph{Gaia} parallaxes will offer a new opportunity to calibrate statistical distance scales such as that of FPB16 using galactic PNe and bring the uncertainties closer to the intrinsic scatter of the relationship. Trigonometric parallaxes provide the most direct means of measuring distances, but their properties mean that they require a proper prior on the underlying distances which must be accounted for at the level of the derived relationship rather than for individual distances such as those published in the catalogue by \citet{bailerjonesdistances}. Selection effects may be present as well as certain types of PNe may be more amenable to distance determination from central star parallaxes. Performing such a calibration is beyond the scope of this particular work, but we believe that the uniform matching performance of our automated technique will offer a good basis for such work in the future, in particular with the improved data in the forthcoming \emph{Gaia} EDR3.

\section{Conclusions} \label{conclusions}

We have used a novel application of the likelihood ratio method to automatically match central stars of planetary nebulae in the HASH PN catalogue with sources in \emph{Gaia} DR2 based on their positions and colours, with a particular focus on accuracy and consistency that contrasts with previous works. Our catalogue of matches includes confidence scores, and is the largest available for \emph{Gaia} DR2 at the time of writing. We have described a few examples of how this catalogue and the new data offered by \emph{Gaia} will enable future science, and discussed the importance of accurate matching in achieving these aims. We emphasise that the certainty of the matching itself should be considered holistically in any analysis.

There are opportunities for further refinement of our matches based on additional data. Photometry from other surveys could disambiguate where \emph{Gaia} colours are lacking, though \emph{Gaia} itself will improve significantly on this front in the future with the full BP/RP spectrophotometry (low resolution spectra). As noted in the previous section, the candidate central star sources with the best parallaxes can be further evaluated on their plausibility as central stars based on their positions in the HR diagram and whether the resulting distance is compatible with the angular size and surface brightness of the nebula itself. Equally, outliers in these parameter spaces can point to interesting sources and systems for followup and further study, such as binary central stars.

Our automated method makes it possible to easily and quickly update the catalogue based on future \emph{Gaia} data releases and future PN discoveries. This will enable us to leverage the improved completeness and more precise astrometric measurements in those future data releases to better understand the galactic PN population.

\begin{acknowledgements}
{We thank the anonymous referee for their comments, which have helped improve this paper.

This research has made use of data from the European Space Agency (ESA) mission {\it Gaia},\footnote{\url{https://www.cosmos.esa.int/gaia}} processed by the {\it Gaia} Data Processing and Analysis Consortium (DPAC).\footnote{\url{https://www.cosmos.esa.int/web/gaia/dpac/consortium}} Funding for the DPAC has been provided by national institutions, in particular the institutions participating in the {\it Gaia} Multilateral Agreement.

This research has also made use of the HASH PN database,\footnote{\url{http://hashpn.space}} of Astropy,\footnote{\url{http://www.astropy.org}} a community-developed core Python package for Astronomy \citep{astropy:2013, astropy:2018}, and of "Aladin sky atlas" developed at CDS, Strasbourg Observatory, France.

Parts of this research were based on data products from observations made with ESO Telescopes at the La Silla Paranal Observatory under programme ID 177.D-3023, as part of the VST Photometric H$\alpha$ Survey of the Southern Galactic Plane and Bulge (VPHAS+),\footnote{\url{www.vphas.eu}} as well as data obtained as part of the INT Photometric H$\alpha$] Survey of the Northern Galactic Plane (IPHAS)\footnote{\url{www.iphas.org}} carried out at the Isaac Newton Telescope (INT). The INT is operated on the island of La Palma by the Isaac Newton Group in the Spanish Observatorio del Roque de los Muchachos of the Instituto de Astrofisica de Canarias. All IPHAS data are processed by the Cambridge Astronomical Survey Unit (CASU), at the Institute of Astronomy in Cambridge.

This research was supported through the Cancer Research UK grant A24042.}
\end{acknowledgements}

\bibliographystyle{aa} 
\bibliography{bib.bib} 

\begin{appendix}
\section{Implementation details}

The probability densities functions (PDFs) used to calculate the likelihood ratio (Eq. \ref{eq:LR}) are estimated from the data themselves. The particular methods and parameters were chosen with the overall aim of producing smooth density ratios with few extrema and to thereby avoid overfitting.

\subsection{Colour density ratio estimation} \label{impdet-cdrest}

Our goal is to estimate the density in BP -- RP colour space of true CSPNe and of non-CSPNe (background sources). We determine our estimates empirically by choosing representative examples of both kinds of sources based on their positions.

The BP and RP fluxes measured by \emph{Gaia} can be contaminated by light from nearby sources (within a couple of arcseconds), particularly in densely populated or nebulous regions. For well-behaved sources with no contamination, it is expected that the total flux measured in the BP and RP passbands should approximately match that of the G band, which does not have the same possibility of contamination. Deviations from this relation are indicated in the catalogue by a large photometric excess factors, and \citet{gaiaphotometry} suggests using cuts based on this factor to select photometrically well-behaved sources for applications relying on colour information. Rather than ignoring the colours of these high excess factor sources completely with hard cuts, we incorporate the excess factor into our density estimation, treating the source colour space as two-dimensional.

We bin sources by excess excess factor (distance above the locus of well-behaved colours, that is $\texttt{phot\_bp\_rp\_excess\_factor}-1.3\times\texttt{bp\_rp}^2$, taken from \citet{gaiaphotometry}) with overlapping bins. We compute the density ratio within each bin as a function of BP -- RP alone, and then smoothly interpolate to get the density ratio values for excess factors between bin centres (interpolating towards one for high excess, corresponding the colour density ratio of one for sources lacking colours). Thus while the density ratio function has a two-dimensional domain, colour densities are only ever one-dimensional in BP -- RP. We thereby hope to treat excess factor as only a quality indicator.

We estimate the density ratios at each BP -- RP value within a single bin non-parametrically using kernel density estimation with a Gaussian kernel. Because of the highly varying density, we use a balloon estimator, in which the kernel bandwidth (the standard deviation of the Gaussian in this case) is variable and is inversely proportional to the local density at the sample point. We estimate the local density from the distance to the $n$th nearest neighbour in BP -- RP, so the kernel bandwidth is effectively proportional to distance to the $n$th nearest neighbour. The bandwidth is clipped to lie within a range of well-behaved values. To avoid artefacts from mismatched kernel widths, the same kernel width is used for both the numerator and denominator of the density ratio, with the kernel chosen based on the numerator density (the density of BP -- RP colours for candidate central stars chosen based on distance or nearest neighbour), since there are fewer such sources.

Sources used to estimate the background colour density (either non-nearest neighbour sources in the first iteration or sources with low separation density ratios in the second iteration) are weighted by the inverse of the local spatial source density $\rho$. The idea of this is that each PN neighbourhood is given equal weighting in the denominator of the colour density ratio estimation (the colour density for background sources). Each PN neighbourhood is by default equally weighted in the estimate for genuine match colours (the numerator), since all neighbourhoods contribute (at most) a single candidate genuine match and are thus weighted equally in that calculation.

\subsection{Separation density ratio estimation}

The set of sources used to estimate the separation density ratio is those with a colour likelihood ratio $> 20$ from the initial (nearest neighbour) colour density ratio estimation. We apply a cutoff on the separation $s$ to these sources, requiring that $s < r_{PN} + 2"$ where $r_{PN}$ is the PN radius in arcseconds, with the addition of $2"$ reflecting our expectation that the relative positional uncertainty is greater for smaller PNe. There are $n$ \emph{Gaia} sources that met our cutoff, having separations $s_i,\ i=1\dots n$. These sources are associated with confirmed ($\texttt{PNstat} = \texttt{T}$ in HASH) PNe with radii < 600\arcsec (including unresolved PNe with no size information in HASH, which we treat for the purposes of binning as having radii of 0.25\arcsec).

As noted in Sect. \ref{positionaluncertainty} and Fig. \ref{fig:separationhistogram}, the distribution of separations $s$ does not match well with a single Rayleigh distribution, unsurprising given the multiple sources of positional uncertainty. However adopting a fully non-parametric approach does not work as well as it did for the colour density in the previous section.

The PDF of a Rayleigh distribution is
\begin{equation}
    \textup{Rayleigh}(r; \sigma) = \frac{r}{\sigma^2}e^{-r^2 / 2\sigma^2}
\end{equation}
which has the convenient property that the $r$ term cancels with the $r$ in the PDF for a constant density of background sources, that is $2\pi r \rho$, giving a likelihood ratio that levels off at a finite value as the separation approaches 0. This reflects the fact that while finding a background source with a very small separation is highly unlikely, so is finding a true counterpart source.

To preserve these properties we form our distribution as by mixing $n$ Rayleigh distributions
\begin{equation}
    f(r) = \sum_{i=1}^n w_i\ \textup{Rayleigh}(r; \sigma_i)
\end{equation}
with parameters $\sigma_i$ each corresponding to the maximum likelihood estimates (MLEs) from a single separation $s_i$, that is $\sigma_i=s_i/{\sqrt 2}$. This mixture captures the behaviour of the empirical distribution while ensuring that the resulting density ratio is smooth, strictly decreasing, and well behaved near zero.

Another advantage of this mixture approach is that the mixture can be reweighted to fit different PN sizes, reflecting the expected dependence in positional uncertainty on the size of the PN. Rather than identical weights $w_i=1/n$, we choose mixture weights for a PN with radius $r_{PN}$ as
\begin{equation}
    w_i \propto \exp\frac{(\log_2 r_{PN} - \log_2 r_{PN_i})^2}{2\sigma^2}
\end{equation}
scaled so that $\sum_i{w_i}=1$. We consider log radii as the logarithm is scale invariant, and choose a standard deviation $\sigma=0.5$ so that most of the influence comes from PN with radii $r_{PN_i}$ within a factor of two of the given PN radius.

\subsection{Justification of nearest neighbour approximation}

We form our initial estimate of the colour density ratio by splitting our candidate set into nearest and non-nearest neighbours, and use the candidate points with the highest colour density ratio as a kind of initial training set for learning the positional uncertainties. This density estimation (and indeed the second iteration based on position) is contaminated in both directions, with many background sources in the nearest neighbour set (standing in for the CSPNe set) and some true CSPNe in the non-nearest neighbour set (standing in for the background distribution).

The effect of this contamination is to push the density ratio towards one (the density ratio becomes one in the limit where the two distributions contain the same proportions of true CSPNe and background sources). We can still learn useful and valid information from the colour provided that true CSPNe make up a larger proportion of the nearest neighbour set than they do of the non-nearest neighbour set, which we expect will be the case as the non-nearest neighbour set is so much larger to begin with.

\end{appendix}
\end{document}